# Data-driven reconstruction of band dispersion and quantum geometry via Koopman dynamic mode decomposition

Yiming Pan[1*], Jinze He[1], Jiapeng Yang[1], Zhiwei Fan[2]

[1]*State Key Laboratory of Quantum Functional Materials, School of Physical Science and Technology and Center for Transformative Science, ShanghaiTech University, Shanghai 200031, China*

[2]*School of Engineering, Newcastle University, Newcastle upon Tyne, NE1 7RU, UK*

## Abstract

We present a data-driven framework for reconstructing band structures using Koopman operator analysis and dynamic mode decomposition (Koopman-DMD). Instead of deriving spectra from an explicit Hamiltonian, the approach reconstructs band dispersion and modal dynamics directly from spatiotemporal data, including wavefunctions and observables. This framework establishes a correspondence between Hamiltonian Floquet-Bloch decomposition and Koopman-DMD, whereby the extracted DMD modes encode frequencies, decay or growth rates, spatial profiles and projection weights. These quantities allow the reconstruction of spectral functions, local density of states, and delocalized-to-localized measures such as the inverse participation ratio. Also, these extended DMD modes enable inference of quantum-geometric and topological properties, including the quantum metric, Berry curvature and geometric phases. Applications to prototypical one- and two-dimensional tight-binding models, including disordered Su-Schrieffer-Heeger model and its Floquet and non-Hermitian variants, graphene and Haldane models, demonstrate that Koopman-DMD provides a unified route for the data-driven analysis of wave propagation, localization, and topological phases in condensed matter, photonics, and related fields.

## Introduction

Band structure engineering seeks to manipulate the spatial or spatiotemporal periodicity of materials so as to tailor the energy ranges (bands) in which electrons in solids [1,2] or photons in dielectrics [3] are allowed or forbidden to propagate. By designing periodicity and material composition, one can create targeted band dispersion and bandgaps that govern electrical transport, optical absorption and emission, and light propagation. In solid-state physics, for example, band-structure engineering modifies semiconductor bandgaps to regulate conductivity and optical response, enabling devices such as transistors, laser diodes, and high-efficiency solar cells. In photonic crystals (PhCs), it controls light propagation and confinement through periodic dielectric structures [4,5], producing photonic bandgaps and engineered dispersion for applications such as high-reflectivity mirrors, PhC fibers and optical cavities.

From a theoretical perspective, systems with spatial or temporal periodicity are conventionally described within the Floquet–Bloch framework [6,7]. The Floquet–Bloch theorem is widely used to compute band structure and identify band modes, defect states within gaps, and group-velocity dispersion (GVD) in periodic media. Here we focus on electronic and photonic systems, for which the band structure – that is, the energy-momentum relation $\varepsilon_n(k)$ – is central to understanding the properties of solids and dielectrics. Conventional band-structure engineering follows an equation-based paradigm (Fig. 1): one first constructs a Hamiltonian $H$ from crystal structure and symmetry, and then solves the corresponding Schrödinger (or Maxwell) eigenvalue problem. In practice, however, this strategy faces several limitations. In many systems, including strongly-correlated electron systems and nonlinear photonic crystals, the exact Hamiltonian may be unknown or only partially constrained. Disorder and defects further break translational symmetry, limiting momentum-space analysis and often necessitating computationally expensive real-space supercell calculations. In addition, periodically driven (Floquet) systems require a quasienergy band description of nonequilibrium dynamics [8–10], which further increases theoretical and numerical complexity.

To address these challenges, data-driven approaches have been increasingly important in modern physics. Rather than relying on prior knowledge of governing equations [11], this paradigm starts directly from spatiotemporal data generated by experiments or simulations to uncover latent physical structure. While techniques such as physics-informed neural networks or symbolic regression excel at discovering explicit governing equations, Koopman operator theory [12–15] and dynamic mode decomposition (DMD) [15,16] provide a particularly natural, equation-free bridge between data and physics by linearly embedding the nonlinear dynamics [17].

The conventional equation-based Floquet–Bloch decomposition can, in principle, be complemented by a data-driven Koopman–DMD decomposition performed directly on spatiotemporal observables. Without prior knowledge of the equations of motion or the detailed system parameters, Koopman–DMD enables the extraction of Floquet–Bloch-like extended modes, as well as localized modes, from state data. The extracted DMD modes can then be used to analyze localization transitions, band topology, and quantum geometry, including Chern numbers, topological edge states, Berry curvature, and the quantum metric. In the present work, this reconstruction strategy is validated on controlled single-particle benchmark models. More broadly, the method is most naturally suited to data-first settings in disordered, driven, or experimentally inaccessible systems where informative spatiotemporal data are available.

In this paper, we establish a theoretical connection between Koopman–DMD analysis and band structures obtained from the Floquet–Bloch formalism, and support this connection with numerical demonstrations in benchmark one- and two-dimensional lattice models. The central question is whether spatiotemporal data can be organized into effective modal excitations, including Bloch-like modes, localized modes, and collective oscillatory structures. Within the data resolution and modal coverage considered here, the benchmark results support this interpretation. The representative models considered here include the SSH model and its variants, graphene, and the Haldane model. Across these representative lattice models, the extracted DMD modes agree closely with those obtained from exact Hamiltonian Floquet–Bloch analysis, thereby validating the reconstruction protocol against known Hamiltonian reference results.

**Koopman DMD versus Hamiltonian FBD**

We present a comparative formulation of Hamiltonian Floquet-Bloch decomposition (FBD) and Koopman dynamic mode decomposition (DMD)**.** Their primary difference is methodological: Hamiltonian FBD is equation-based, whereas Koopman DMD is equation-free but data-driven**.** Equivalently, the two approaches differ most directly at the input level. Hamiltonian FBD starts from an explicit or calibrated Hamiltonian, such as the Schrödinger equation, Maxwell's equations, tight-binding models, or coupled-mode equations. Koopman-DMD starts instead from spatiotemporal data, such as wavefunctions, electromagnetic fields, density/current fields, correlation functions, or expectation-value time traces, and constructs an effective evolution operator from the snapshots. Here we focus on crystalline solid-state and photonic lattice systems, where these two routes can be compared directly using controlled benchmark models. By construction, Koopman-DMD is model-agnostic and equation-free, but it places stringent demands on the quality and richness of spatiotemporal data. When the data are sparse, noisy, or poorly resolved, or insufficiently excited over the relevant modal sectors, the extracted DMD modes become unreliable. Appropriate sampling conditions and SVD truncation can reduce numerical noise and artifacts, while different initial states or datasets can improve modal coverage. We assess robustness against readout noise using SSH-model snapshots with added random noise. This analysis shows how truncation balances noise suppression against the retention of physical modes, and reveals different noise sensitivities in the reconstruction of band dispersion, quantum metric, and Zak phase (SM, Sec. S3 and Figs. S4–S7). Disorder modifies the underlying system dynamics and should be distinguished from readout noise, which directly perturbs the measured snapshots. This requirement on data quality reflects a fundamental difference between the two approaches. In Hamiltonian FBD, physical insight depends on the accuracy of the model Hamiltonian and governing equations, whereas in Koopman-DMD, it is encoded in the dataset itself. Issues of data acquisition, cleaning, classification, and preprocessing are discussed in the Supplementary Material (SM).

**Basic DMD framework**. DMD approximates complex spatiotemporal data by decomposing them into spatial modes, modal amplitudes, and temporal dynamics. Given a sequence of snapshots $x(t_k)$, we form the data matrix $X = [x(t1)\ x(t2)\ \cdots\ x(tm)]$, and seek a low rank-r approximation of the form $X \approx \Phi\ diag(w)T$, where $\Phi = [\phi_1, \dots, \phi_r]$ contains the spatial DMD modes, $w = (w_1, \dots, w_r)^T$ gives the projection weights and the temporal matrix $T$ is built from exponentials $e^{i\omega_i t}$. In this decomposition, the spatial modes $\phi_j$ represent coherent structures in space, including extended waves with characteristic momenta and localized modes with finite confinement lengths, where the complex frequencies $\omega_j$ determine the temporal evolution of each mode, including oscillation and decay or growth.

Accordingly,

$$X \approx \begin{bmatrix} | & | & | \\ \phi_1 & \cdots & \phi_r \\ | & | & | \end{bmatrix} \begin{bmatrix} w_1 & & \\ & \ddots & \\ & & w_r \end{bmatrix} \begin{bmatrix} e^{-i\omega_1 t_1} & \cdots & e^{-i\omega_1 t_m} \\ \vdots & \ddots & \vdots \\ e^{-i\omega_r t_1} & \cdots & e^{-i\omega_r t_m} \end{bmatrix}, \tag{1}$$

so that the data can be written as

$$\mathrm{x(t)} \approx \sum_{j=1}^{r} \phi_j w_j e^{-i\omega_j t} = \underbrace{\sum_{i}^{\prime} u_i(k) w_i e^{-i\omega_i t + i\boldsymbol{k}_i \cdot \boldsymbol{r}}}_{\text{Extended modes}} + \underbrace{\sum_{j}^{\prime\prime} \phi_j w_j e^{-i\omega_j t}}_{\text{Localized modes}} . \tag{2}$$

We broadly classify the DMD modes into extended and localized branches, which will be later quantified using the inverse participation ratio (IPR) [18–20]. For Bloch-like modes with $\phi_i = u_i(k)e^{i\boldsymbol{k}_i \cdot \boldsymbol{r}}$, the relation between the dominant momentum $k_i$ and the oscillation frequency $\Re\{\omega_i\}$ gives the band structure $\omega = \omega(\boldsymbol{k})$, while $\Im\{\omega_i\}$ characterizes growth or decay. In the applications considered here, the snapshot $x(t)$ may represent the wavefunction $\psi(x,t)$ in a quantum system, or the electric field $E(x,t)$ in an electromagnetic system, in either discrete or continuous spacetime. Thus, for sufficiently resolved state-level data, the DMD frequency, spatial mode, and projection weight play roles analogous to the quasienergy, Floquet-Bloch mode, and modal amplitude in Hamiltonian FBD.

**Basic FBD framework.** The Floquet–Bloch theorem provides a standard framework for systems with spatial and temporal periodicity. For a periodically driven quantum system with $H(\boldsymbol{r}+\boldsymbol{R},t) = H(\boldsymbol{r},t)$ and $H(\boldsymbol{r},t+T) = H(\boldsymbol{r},t)$, where $\boldsymbol{R}$ is a lattice vector and $T$ is the driving period, the wavefunction can be written in Bloch form, $\psi_{n,\boldsymbol{k}}(\boldsymbol{r},t) = e^{i\boldsymbol{k}\cdot\boldsymbol{r}} u_{n,\boldsymbol{k}}(\boldsymbol{r},t)$, where $u_{n,\boldsymbol{k}}$ is lattice-periodic in space. Temporal periodicity further implies, through Floquet theory, that $|\psi_\alpha(t)\rangle = e^{-i\varepsilon_\alpha t}|u_\alpha(t)\rangle, |u_\alpha(t)\rangle = |u_\alpha(t+T)\rangle$. Expanding both the Hamiltonian and the Floquet mode in Fourier harmonics converts the time-dependent Schrödinger equation into an eigenvalue problem, whose truncated solution yields the quasiband structure $\varepsilon_\alpha(\boldsymbol{k})$ and the corresponding Floquet–Bloch modes.

Once the eigenvalues and eigenstates are obtained, the Hamiltonian framework can be used to construct several classes of physical observables. Firstly, the band structure determines dynamical properties such as the energy gap, group velocity, and group-velocity dispersion, which controls pulse broadening and nonlinear propagation. In systems with boundaries, defects, or disorder, the same framework also yields localized states such as edge and defect modes. Secondly, quantum geometric and topological properties follow from eigenstates across the Brillouin zone. The Berry connection is $A_{nm}(k) = i\langle u_{n,k} | \nabla_k | u_{m,k} \rangle$ [21], the Berry phase is obtained by integrating the diagonal Berry connection along a closed loop in momentum space, and quantum geometric tensor can be further defined accordingly [22–26]. Thirdly, for interacting or many-body systems, one often introduces Green's functions within a mean-field or self-energy framework, $G(\omega,k) = (\omega - H(k) - \Sigma(\omega,k))^{-1}$ from which the spectral function, local density of states, and quasiparticle lifetime can be derived. For example, $A(\omega, k) = -\frac{1}{\pi}\mathrm{Im}G(\omega, k)$ up to the convention used.

Despite its rigor, the Hamiltonian FBD approach has a clear limitation: it requires an explicit and sufficiently accurate model Hamiltonian. For disordered, strongly nonlinear or experimentally inaccessible systems, constructing such a Hamiltonian may be impractical or impossible. Even when an effective Hamiltonian can be proposed, practical modeling often involves simulating the dynamics, comparing it with observed spatiotemporal data, and adjusting model parameters before the Hamiltonian-based band analysis is performed. This limitation motivates the Koopman–DMD approach, which shifts the starting point from equations to data.

**Koopman linearization.** Koopman operator theory has emerged as a powerful framework in fluid dynamics, control theory, pure mathematics, and, more recently, physics [17,27,28]. Its central idea is to lift a finite-dimensional nonlinear dynamical system into a linear evolution in an infinite-dimensional observable space. For a discrete-time dynamical system $x_{t+1} = F(x_t), x \in M$, and an observable $g: M \to C$, the Koopman operator $\mathcal{K}$ acts as $\mathcal{K}g(x_t) = g\big(F(x_t)\big) = g(x_{t+1})$. Although the map $F$ may be nonlinear, $\mathcal{K}$ is linear and one may therefore consider its spectral decomposition, $\mathcal{K}\varphi_j(x) = \lambda_j\varphi_j(x)$. Here $\lambda_j$ are Koopman eigenvalues and $\varphi_j$ are Koopman eigenfunctions. Physically, the eigenvalues encode oscillation frequencies, whereas the corresponding modes represent spatiotemporal coherent structures analogous to Floquet–Bloch eigenmodes.

Dynamic mode decomposition (DMD) provides a practical finite-data realization of Koopman spectral analysis. Given a sequence of spatiotemporal snapshots, such as wavefunctions $\psi(x, t_n)$, DMD seeks a best-fit linear evolution operator $A$ satisfying $\psi(t_{n+1}) \approx A\,\psi(t_n)$. Using singular value decomposition and low-rank regression [13–16], one obtains the eigenvalues $\lambda_j$ and eigenmodes $\phi_j(x)$ of $A$. Writing $\lambda_j = e^{-(\gamma_j + i\omega_j)\Delta t}$, the modal frequency and decay rate are $\omega_j = -\mathrm{Im}\big[\ln\lambda_j\big]/\Delta t$, $\gamma_j = -\mathrm{Re}\big[\ln\lambda_j\big]/\Delta t$, where $\Delta t$ is the sampling interval. In conservative systems, $\gamma_j \approx 0$, so the eigenvalues lie near the unit circle and $\omega_j$ gives the intrinsic modal frequency. In non-Hermitian or dissipative systems, nonzero $\gamma_j$ measures modal amplification or decay.

The spatial profile $\phi_j(x)$ gives the corresponding mode shape in real or configuration space. For extended modes, Fourier analysis of $\phi_j(x)$ allows one to extract a dominant momentum label $k$. For localized modes, by contrast, a sharp momentum is generally ill-defined, and localization is better characterized by a localization length or inverse participation ratio (IPR). This distinction also clarifies the relation between Koopman-DMD and conventional Fourier analysis. Fourier analysis is powerful for identifying frequency–momentum spectral content, while Koopman-DMD further separates the data into coherent modes with complex frequencies, decay or growth rates, projection weights, and real-space profiles. In this way, DMD extracts the frequency spectrum, decay (or growth) rate, spatial mode profile, and projection weight directly from data, without prior knowledge of the underlying governing equations. This makes it particularly useful for inferring band structure and its associated dynamics from simulations or experiments.

**Schrodinger picture in state space versus Heisenberg picture in observable space.** There is a formal equivalence between analyzing the dynamics of wavefunctions and those of observables, although the two descriptions differ in the information that is directly accessible in practice. In the Heisenberg picture, an observable $O(t)$ evolves according to $dO/dt = \frac{i}{\hbar}[H, O]$, just as the Schrödinger picture evolves quantum states. Formally, this evolution may be written as:

$$O(t) = e^{\mathcal{L}t}O(0), \tag{3}$$

where $\mathcal{L}[O] = \frac{i}{\hbar}[H, O]$ is the Liouville superoperator. This structure is analogous to the action of the Koopman operator on observables. The two pictures are equivalent and describe the same dynamics in different representations. Crucially, the linear evolution of observables provides a natural foundation for Koopman-based DMD, because observable dynamics can be embedded into a linear operator framework.

Correspondingly, Koopman-DMD can also be applied to a sequence of observables: $O = \{\hat{O}_1, \hat{O}_2, \cdots, \hat{O}_n\}^T$. When only truncated or local limited measurements are available, we construct delay-embedded Hankel matrices [29],

$$\mathrm{H}_0 = \begin{bmatrix} \hat{O}_0 & \hat{O}_1 & \cdots & \hat{O}_{m-1} \\ \hat{O}_1 & \hat{O}_2 & \cdots & \hat{O}_m \\ \vdots & \vdots & \ddots & \vdots \\ \hat{O}_{q-1} & \hat{O}_q & \cdots & \hat{O}_{q+m-1} \end{bmatrix}, \mathrm{H}_1 = \begin{bmatrix} \hat{O}_1 & \hat{O}_2 & \cdots & \hat{O}_m \\ \hat{O}_2 & \hat{O}_3 & \cdots & \hat{O}_{1+m} \\ \vdots & \vdots & \ddots & \vdots \\ \hat{O}_q & \hat{O}_{q+1} & \cdots & \hat{O}_{q+m} \end{bmatrix}, \tag{4}$$

and approximate the evolution by

$$\mathrm{H}_1 \approx K\mathrm{H}_0. \tag{5}$$

The effective operator $K$ is obtained from a low-rank regression or optimization procedure [30]. The delay embedding does not remove memory or non-Markovian effects; rather, it absorbs them into an enlarged effective observable (or state) space. In this sense, the unresolved operator hierarchy is encoded in hidden coordinates stored in the delay stack. For instance, please find the detailed discussion of Green's functions [31,32] in the SM file.

If the full wavefunction $\Psi(x, t)$, or equivalently the state vector, is accessible in time, DMD provides a particularly direct route to the system's eigenstates and eigenenergies. Writing $\Psi(t) = \sum_n c_n \phi_n e^{-iE_n t/\hbar}$, one expects DMD to identify modes with frequencies $\omega_n = E_n/\hbar$ together with spatial profiles $\phi_n(x)$ corresponding to the underlying eigenfunctions, such as Bloch waves or localized states. This is the ideal setting for DMD because the data already lie in the linear subspace spanned by the true eigenmodes. In this case, the Koopman operator coincides with the unitary evolution generated by the Hamiltonian, and DMD recovers its spectral decomposition.

In practice, one usually has access only to observables, such as densities, currents, or limited expectation values, rather than to the full wavefunction. Their temporal evolution is therefore more indirect. For a generic observable $O(t)$, the measured signal is determined by matrix elements between levels and thus oscillates at optical frequencies, that is, energy differences $(E_m - E_n)/\hbar$, rather than at the absolute eigenenergies themselves. If $O_{mn} = \langle m|O|n\rangle$, then for a superposition of eigenstates the observable contains terms of the form

$$O_{mn} e^{i(E_m - E_n)t/\hbar}. \quad (6)$$

Accordingly, DMD applied to one-time observable data typically identifies transition frequencies rather than absolute single-particle eigenenergies. Recovering absolute eigenenergies generally requires additional structure, such as specially prepared initial states, multiple measurement channels, or two-time correlation functions that provide an effective reference.

Despite this distinction, the Schrödinger and Heisenberg descriptions remain theoretically consistent. If one has access to a complete basis of observables spanning Koopman operator space, then analyzing observable dynamics would be equivalent to analyzing the full state dynamics, since both are governed by the underlying physics. The practical difference is that wavefunction-based DMD is more direct and more information-complete, whereas observable-based DMD must contend with incomplete measurements and possible loss of phase information. As a result, observable-driven DMD often requires multiple initial conditions or repeated measurements to reconstruct the full band structure and its topology reliably. Numerically, DMD in the Schrödinger picture is conceptually cleaner and provides the most direct access to band physics when full state data are available. DMD in the Heisenberg (or observable) picture is more realistic for experiments and remains fully compatible with the Koopman framework, but its modal content requires greater care in interpretation because it more naturally encodes transition frequencies and collective response rather than the full state spectrum itself.

**Band Physics Extraction from DMD modes.**

**Band dispersion.** Within Hamiltonian Floquet–Bloch decomposition (FBD), band dispersion is obtained by first specifying the Hamiltonian $H$, or $H(t)$ in the periodically driven case. For a static periodic lattice, Bloch's theorem gives $\psi_{n\boldsymbol{k}}(\boldsymbol{r}) = e^{i\boldsymbol{k}\cdot\boldsymbol{r}} u_{n\boldsymbol{k}}(\boldsymbol{r})$, where $\boldsymbol{k}$ is the crystal momentum and $n$ labels the band. Substituting this form into the Schrödinger equation yields the band dispersion $E_n(\boldsymbol{k})$. For periodically driven systems, Floquet theory treats time as an additional periodic degree of freedom and leads to a quasienergy spectrum through the Floquet Hamiltonian or, equivalently, the one-period evolution operator $U(T)$ [9,33]. One thus obtains Floquet–Bloch wavefunctions and quasibands that characterize the effective dispersion of driven systems.

Once the eigenstates are known, FBD provides direct access to band-derived observables. The band gap distinguishes insulating, semiconducting, and metallic behavior; the group velocity follows from $v_g = \boldsymbol{\nabla}_k E_n(\boldsymbol{k})$; and second-order derivatives determine group-velocity dispersion, which governs pulse broadening and light propagation in dielectric media. More generally, band analysis based on the Hamiltonian-based analysis provides the spectral function $A(\omega, \boldsymbol{k})$, the local density of states (LDOS), and quasiparticle lifetimes. In optics, the imaginary part of the eigenfrequency further determines radiative loss, from which the quality factor $Q$ and associated field enhancement can be inferred.

In the Koopman–DMD framework, these quantities are extracted directly from spatiotemporal data rather than from an explicit Hamiltonian. Given sampled wavefunction or state data $\psi(x, \mathrm{t})$, the DMD frequencies $\omega_j$ can be interpreted as dominant modal energies, while the corresponding spatial modes $\phi_j(x)$ encode extended or localized structures. For example, in a one-dimensional SSH chain excited by a random initial pulse, DMD applied to the site-resolved dynamics can recover two dominant modal branches corresponding to the upper and lower bands (Fig. 3a), with their separation reproducing the band gap [33,34]. In the topological phase under open boundary conditions, DMD additionally identifies a mode with $\omega \approx 0$, localized near the chain ends and exponentially decaying into the bulk, consistent with the expected topological edge state (see Fig. 3c, and Fig. S2 SM). Accordingly, DMD provides a data-driven route to detecting topological boundary modes directly from dynamics.

For general two-dimensional systems, if the excitation probes a broad region of momentum space, DMD can still recover an effective dispersion relation from the measured data. In graphene-like systems, the reconstructed dispersion is expected to reproduce the Dirac-cone structure $E \propto |\boldsymbol{k}|$ near the band-touching points (Fig. 6). More broadly, this illustrates how Koopman–DMD can infer bulk dispersion from finite-size, finite-time measurements, thereby bridging experimentally accessible dynamics and ideal Bloch-band theory.

**Topological invariants.** Topological phases arise from global geometric properties of band eigenstates rather than from local energetics alone. In Hamiltonian FBD, these quantities are computed directly from Bloch or Floquet eigenstates. In one-dimension, the Zak phase is

$$\gamma = \int_{\mathrm{BZ}} \mathcal{A}(k)\ dk, \mathcal{A}(k) = i\langle u_k | \partial_k u_k \rangle \tag{7}$$

with the SSH model providing the canonical example: $\gamma_{\mathrm{Zak}} = \pi$ in the topologically nontrivial phase and 0 in the trivial phase. In two dimensions, the corresponding invariant is often the Chern number,

$$C = \frac{1}{2\pi} \iint_{\mathrm{BZ}} \Omega(k_x, k_y)\ dk_x dk_y, \tag{8}$$

where $\Omega = \nabla_{\mathrm{k}} \times \mathcal{A}(k)$ is the Berry curvature. Although graphene has zero total Chern number with time-reversal symmetry breaking, each Dirac valley still carries a nontrivial Berry phase, underlying its quantum Hall effect. We will discuss the band topology of Haldane model [35] and graphene [36] in the following sections.

Inferring band topology from DMD modes is less direct, because topological invariants are global quantities defined over momentum or parameter space. Nevertheless, several data-driven strategies are available. One is to reconstruct the evolution of modes under parameter scanning, for example by varying the dimerization ratio $v/w$ in the SSH model and tracking the phase evolution of the extracted DMD modes. Another practical approach is to infer topology from dynamical observables whose long-time behavior is known to reflect topological invariants. In one-dimensional chiral systems, for instance, the mean chiral displacement converges to the Zak phase and thus provides a dynamical signature of topology without direct access to Bloch-state phases. More generally, zero-energy localized edge states or systematic sublattice polarization in the extracted spatial modes can provide strong evidence for nontrivial bulk topology through bulk–boundary correspondence.

For a sufficiently smooth family of DMD bulk modes $|u^{DMD}(k_j)\rangle$, discrete Berry phases can be computed directly from overlap products. For instance, the Zak phase can be evaluated using the gauge-invariant discretized formula

$$\gamma_{Zak} = -\text{Im}\ \ln\prod_{j=1}^{N}\left\langle u_{k_j}^{DMD}\,\middle|\,u_{k_{j+1}}^{DMD}\right\rangle, \tag{9}$$

Similarly, the Berry flux through each plaquette is obtained from Wilson-loop overlap products, and summation over the discretized Brillouin zone yields the Chern number. This can be implemented using the Fukui–Hatsugai–Suzuki (FHS) construction [37], in which the link variable

$$U_\delta(\mathrm{k}) = \frac{\langle\, u(k)\mid u(k+\delta)\,\rangle}{|\langle\, u(k)\mid u(k+\delta)\,\rangle|}, \tag{10}$$

removes the arbitrary $U(1)$ phase of independently extracted DMD modes. In this way, the computation of Chern numbers or winding numbers from DMD modes becomes formally parallel to that in Hamiltonian FBD, while retaining sensitivity to boundaries, defects, and finite-size structure.

**Berry curvature and quantum metric.** Beyond topological invariants, the geometric structure of a band is encoded in the quantum geometric tensor (QGT) [22,38], whose imaginary part gives the Berry curvature ($\Omega_{\mu\nu}$) and whose real part gives the quantum metric ($g_{\mu\nu}$). The Berry curvature governs topological phenomena such as Quantum Hall Effect [39,40], whereas the quantum metric measures the distance between neighboring Floquet-Bloch states. Reconstructing the QGT therefore provides direct access to the quantum geometry of the extracted DMD modes in momentum space.

For instance, the real part of the QGT, namely the quantum metric $g_{\mu\nu}$, measures the Fubini-Study distance between nearby quantum states. Rather than evaluating derivatives explicitly, the discrete quantum metric can be extracted from the fidelity (inner product overlap) between adjacent DMD modes:

$$ds^2 = g_{\mu\nu}dk^\mu dk^\nu \approx 1 - \left|\langle u^{DMD}(k_i)|u^{DMD}(k_j)\rangle\right|^2. \tag{11}$$

Explicit expressions for $g_{xx}, g_{yy}, g_{xy}$ are given in the SM file. This formulation is insensitive to local phase fluctuations because it depends only on gauge-invariant overlaps. In a discretized DMD framework, the local Berry curvature on a momentum-space plaquette centered at k is

$$\Omega_{xy}(k) = \text{Im}\ln\left[U_x(k)U_y(k+\delta k_x)U_x^*(k+\delta k_y)U_y^*(k)\right]/\left(\delta k_x\delta k_y\right). \tag{12}$$

The diagonal components of the quantum metric can be estimated from the norms of the link variables, whereas off-diagonal components follow from cross-overlap terms. This fully discrete FHS formulation is gauge invariant and is therefore well suited to Koopman-DMD framework, where each extracted DMD mode carries an arbitrary phase choice. It enables reconstruction not only of topological invariants but also of the quantum geometric tensors of the data-driven band structure.

**Anderson localization and phase transition.** In lattice systems, increasing disorder strength W drives the system toward an Anderson-insulating regime [41,42]. Once disorder breaks translational symmetry, crystal momentum k ceases to be a good quantum number, and conventional Hamiltonian FBD is no longer directly applicable. Koopman–DMD then provides a natural alternative by operating on real-space dynamical data. Applying DMD to spatiotemporal wavefunction yields spatial modes $\phi_j(x)$ that, in the localized regime, exhibit exponential envelopes,

$$\left|\phi_j(x)\right| \sim e^{-|x-x_\mathrm{j}|/\xi_\mathrm{j}}, \tag{13}$$

from which the localization length $\xi_\mathrm{j}$ can be estimated by envelope fitting or, more robustly, by through inverse participation ratio (IPR) [18]. The disappearance of extended modes and the emergence of localized modes in the DMD spectrum therefore provide a direct dynamical signature of the localization-to-delocalization Anderson transition (see Fig.2).

From the Hamiltonian perspective, disorder invalidates Bloch reduction and generally requires large-scale exact diagonalization, real-space numerical methods, or alternative topological diagnostics such as the Bott index [43]. Topological edge states are often robust against weak disorder as long as the band gap remains open, but their computation typically becomes fully numerical rather than analytic. By contrast, the Koopman–DMD framework does not rely on the strict periodicity and is therefore particularly well suited to finite disordered samples. In a 1D disordered system, even arbitrarily weak disorder localizes all eigenstates. Although Bloch theory fails, the dynamics remain measurable: a localized initial wavepacket stays confined within a finite spatial range. DMD then resolves discrete localized modes with distinct frequencies, each centered at a different spatial position. In this regime, a Fourier spectrum can still reveal broad frequency or momentum content, but it does not directly isolate individual localized real-space modes. In this sense, DMD plays a role analogous to real-space diagonalization while operating directly on data, and thus provides a practical route to extracting localized modes and their spectrum.

A useful quantitative diagnostic is the inverse participation ratio,

$$IPR = \frac{\Sigma\left|\phi_j\right|^4}{\left(\Sigma\left|\phi_j\right|^2\right)^2} \tag{14}$$

which distinguishes extended and localized states through its scaling with system size N. For extended states, IPR $\sim O(1/N)$, whereas for localized states it tends to a nonzero constant as $N \to \infty$ [18–20]. Applied to DMD modes, the IPR provides a practical measure of real-space mode concentration and hence of localization. The same criterion can be generalized to other settings, including spin chains, and many-body Hilbert or Fock spaces [18].

Disorder can also modify topological phases. In 1D systems such as the SSH chain, if disorder preserves chiral symmetry and does not close the gap, the Zak phase remains well defined, although its computation then requires real-space formulations, such as polarization-based approaches, rather than Bloch-based expressions. Koopman–DMD is particularly useful when disorder creates or suppresses boundary-localized modes. In this sense, DMD provides a data-driven route to detecting topological transitions in finite disordered systems. Its main strength is the direct extraction of edge or defect spectral modes from individual samples, whereas global topological invariants are generally less directly accessible.

Notably, for disordered many-body systems, exact diagonalization is often used to distinguish localized and thermal phases through level-spacing statistics, such as Poisson versus Wigner–Dyson distributions [44]. However, exact diagonalization rapidly becomes computationally expensive as system size grows. For Koopman–DMD, one may instead examine the frequency distribution and spatial support of localized DMD modes extracted from experimentally or numerically accessible observables. Although this does not replace conventional many-body diagnostics, it offers a practical alternative for identifying many-body localization behavior in systems where direct diagonalization is difficult.

**Photonic bandgap diagnostic.** In photonic crystals, Maxwell's equations can be written in the operator form $\Theta H = \frac{\omega^2}{c^2} H, \Theta = \nabla \times \frac{1}{\varepsilon(r)} \nabla$, which is formally analogous to a Hamiltonian eigenvalue problem [3]. In practice, however, the dielectric profile $\varepsilon(r)$ is often not known with sufficient precision because of fabrication disorder or incomplete microscopic characterization. What is directly accessible instead is the spatiotemporal electromagnetic field, for example through near-field scanning optical microscopy (NSOM) or related field-resolved measurements. This makes photonic systems particularly well suited to Koopman–DMD analysis.

In the DMD workflow, one starts from measured field snapshots $E(r, t)$ or $H(r, t)$, decomposes them into modes with distinct frequencies $\omega_j$, and then performs spatial Fourier analysis on the extracted DMD modes to identify their dominant wavevectors k. This yields a reconstructed $\omega$-$k$ dispersion relation. The resulting band structure can then be used for photonic tasks such as optimizing group-velocity dispersion for slow-light management or dispersion compensation. In topological photonic platforms containing defects, interfaces, or domain walls, DMD can further isolate protected edge modes, distinguish them from bulk radiation channels, and probe their robustness against backscattering. In photonic systems, the decay rate extracted from each DMD mode determines the modal linewidth and thus the quality factor, while the associated spatial mode profile yields a diagnostic of field enhancement.

**Koopman-DMD workflow for band dispersion and band topology.**

To make the procedure explicit, we organize the Koopman-DMD workflow as follows. Additional discussions and implementation details are provided in the SM file.

**1. DMD of spatiotemporal data.** Suppose that the input data consist of lattice wavefunction snapshots $\psi_j(t)$, or their discretized continuum analogues. We first construct the snapshot matrices:

$$X_1 = [\psi(t_0), \psi(t_1), \dots, \psi(t_{M-1})], \qquad X_2 = [\psi(t_1), \psi(t_2), \dots, \psi(t_M)],$$

and compute standard, optimized, or low-rank DMD through $X_2 \approx AX_1$. Diagonalization of A yields eigenpairs $(\lambda_l, \phi_l)$, where $\phi_l$ are the DMD spatial modes and $\lambda_l$ are the corresponding discrete-time eigenvalues. These are converted into continuous complex frequencies through

$$\omega_l - i\gamma_l = \frac{i}{\Delta t}\log\lambda_l . \tag{15}$$

Here, $\omega_l$ gives the modal oscillation frequency, while $\gamma_l$ gives the effective decay or gain rate.

For extended modes, the associated spectral peak may be represented by a Lorentzian line shape, $\frac{1}{\pi}\frac{\gamma_l}{(\omega-\omega_l)^2+\gamma_l^2}$, whereas for localized resonant modes one may additionally define a quality factor $Q_l = \omega_l/2|\gamma_l|$. These quantities distinguish Bloch-like modes from cavity-like or other spatially bounded resonances. DMD mode localization is quantified through the inverse participation ratio (IPR)

$$IPR_l = \frac{\sum_j |\phi_l(j)|^4}{(\sum|\phi_l(j)|^2)^2} \tag{16}$$

This measure sharply distinguishes extended and localized regimes. For interacting systems, one may instead define a generalized IPR based on the local density of states, $GIPR(\omega) = \sum_j \rho_j(\omega)^2 / \left(\sum_j \rho_j(\omega)\right)^2$, which has proven useful in localization diagnostics and many-body studies [18].

**3. Band structure, gap, group velocity, and GVD.** To reconstruct the dispersion, each extended DMD mode is projected into momentum space,

$$u_l(k) = \sum_j e^{-ik\cdot r_j}\, \phi_l(j). \tag{17}$$

The dominant frequency branch at each k yields the band relation $E(k) \simeq \hbar\omega(k)$, up to convention. From the reconstructed dispersion one may determine the band gap $\Delta$, the group velocity $v_g = \partial E/\partial k$, and the group-velocity dispersion GVD$= \partial^2 k/\partial\omega^2$. Notably, in solid-state context, the local curvature of the reconstructed band may also be used to define an effective mass $m^*$ for the quasiparticle excitations.

**4. Spectral density and local DOS.** An effective momentum-resolved spectral density may be constructed from the DMD modes as a Lorentzian sum weighted by their momentum-space amplitudes,

$$A_{\mathrm{DMD}}(k,\omega) = \sum_l W_l(k)\, \frac{\gamma_l}{(\omega-\omega_l)^2+\gamma_l^2}, \tag{18}$$

where $W_l(k)$ is the modal projection weight. For wavefunction data, this object should be interpreted as a DMD-inferred dynamical spectral density rather than as the exact quasiparticle spectral function. In noninteracting, or mean-field setting, it plays essentially the same practical role: peak positions determine the dispersion, peak weights reflect modal occupation and projection from the initial state, and peak widths encode effective decay or finite-window broadening. However, this spectral density is not, in general, identical to the strict many-body spectral function $A(k,\omega) = -\frac{1}{\pi}\mathrm{Im}\, G^R(k,\omega)$, and the associated modal weights $W_l(k)$ are not automatically equal to the true quasiparticle residue $Z_k$, unless the input data are themselves Green's functions or related correlation functions. If the goal is instead the true many-body spectral function, the input data should not be ψ(t) itself, but rather the time series of single-particle Green's functions, $G_{ij}(t)$ or $G(k,t)$, to which the same DMD procedure is then applied [31,32]. In that case, the resulting $A(k,\omega)$ is directly aligned with the standard Green's-function definition (see SM file).

Likewise, a DMD analogue of the local density of states (LDOS) is obtained from the real-space modal amplitudes,

$$\rho_j(\omega) = \frac{1}{\pi}\sum_l |\phi_l(j)|^2 \frac{\gamma_l}{(\omega-\omega_l)^2+\gamma_l^2} \tag{19}$$

which plays the role of an STM-like local spectral density.

**5. Topological invariants and quantum geometric tensors.** Topological phases may be computed from a smooth family of DMD bulk modes using the same discrete gauge-invariant formulas used in Hamiltonian band theory. In 1D, the Zak phase is evaluated as

$$\gamma_{Zak} = -Im \ln\left(\prod_j \left\langle u_{k_j}\middle|u_{k_{j+1}}\right\rangle\right). \tag{20}$$

In 2D, the Chern number is obtained from the Berry flux on each plaquette

$$C = \frac{1}{2\pi}\sum_{\text{plaquettes}} \mathrm{Im}\ln\big(\langle u_k|u_{k+\delta x}\rangle\langle u_{k+\delta x}|u_{k+\delta x+\delta y}\rangle\langle u_{k+\delta x+\delta y}|u_{k+\delta y}\rangle\langle u_{k+\delta y}|u_k\rangle\big), \tag{21}$$

with gauge ambiguity handled through the Fukui–Hatsugai–Suzuki construction [16]. The same framework also allows extraction of the discrete Berry curvature ($\Omega_{xy}$) and the quantum metric ($g_{xx}, g_{yy}, g_{xy}$) from DMD modes, as described in the SM file.

**Benchmark Results and Discussions**

We consider several representative tight-binding-approximated (TBA) models to investigate the localization and topological behavior in one- and two-dimensional lattice systems. Using the Dynamic Mode Decomposition (DMD) framework, we systematically extract band structures, spatial modal profiles, and topological invariants purely from spatiotemporal evolution data, and validate the extracted results by comparison with Hamiltonian-based reference calculations. The explicit Hamiltonians of these TBA models and their conventional band analysis are given in the SM file. We begin our DMD analysis with 1D models: the Anderson model [45], and the Su-Schrieffer-Heeger (SSH) model [46].

**Localization in the Anderson Model.** The 1D Anderson model describes non-interacting particles in a periodic lattice subjected to random, uncorrelated on-site potentials. According to the scaling theory of localization, any disorder in a purely 1D system leads to the exponential localization of all single-particle eigenstates. To visualize this, we apply DMD to extract the intrinsic spectral and spatial properties from the spatiotemporal snapshots (Fig. 2a). The resulting reconstruction of the band structure and spectral weights shows excellent agreement with theoretical results (Fig. 2b).

By evaluating the spatial profiles and the inverse participation ratio (IPR) of the DMD-extracted modes, we clearly observe the crossover from spatially extended states in the weak-disorder regime (Fig. 2c) to exponentially localized states in the strong-disorder regime (Fig. 2d). This localization transition is further quantified by the modal distribution in the energy–IPR plane (Fig. 2e) and the monotonically decreasing localization length as disorder increases (Fig. 2f). Furthermore, from level-spacing statistics of the DMD modes, we find that the weak-disorder regime (W=0.8) agrees well with the Wigner–Dyson distribution, whereas the strong-disorder regime (W=8.0) agrees well with the Poisson distribution [44], thereby statistically capturing the crossover from extended to localized states (Fig. 2g, h). These results demonstrate that the data-driven DMD approach not only captures the spatial confinement but also resolves the deep, underlying spectral statistics of Anderson transition. In addition, discussions of the Aubry-André-Harper (AAH) model and its localization and topological behavior [47,48] is provided in the SM file.

**Topological-to-Anderson Transitions in SSH Models** The static Su-Schrieffer-Heeger (SSH) model hosts topological zero-energy modes, while periodic driving in the Floquet setting generates anomalous topological phases featuring robust 0- and π -modes [49,50]. Introducing random on-site disorder to these systems induces a competition between topological localization and Anderson localization.

For the static disordered SSH model, we validate the DMD approach by comparing the gapped spectral functions (Fig. 3a) with that result for Anderson model (Fig. 2b). As the disorder strength $W$ increases, the extracted spectrum shows that the initially isolated topological zero modes gradually merge into the Anderson regime (Fig. 3b), as reflected by the overall increase in the IPR with $W$. The spatial profiles further confirm the edge localization of the topological modes (Fig. 3c). Notably, the Anderson transition is also reflected in the IPR mapping of the edge and bulk states: for small disorder ($W = 0.1$), the IPRs are different from each other, but for large disorder ($W = 2$), the two distributions merge (Fig. 3d).

Our DMD framework extend naturally to periodically driven systems by directly processing the time-domain stroboscopic snapshots, without constructing the explicit Floquet Hamiltonian (Fig. 3e). For the disordered driven SSH model, DMD reconstructs the quasienergy as a function of $W$ with the IPR shown as a color scale, and clearly resolves the coexistence of 0- and $\pi$-modes at weak disorder (Fig. 3f). The corresponding spatial profiles confirm their topological localization at boundaries (Fig. 3g). As $W$ increases, these Floquet topological modes are gradually overwhelmed into the Anderson-localized regime, accompanied by a corresponding redistribution of the IPR (Fig. 3h). These results show that the Koopman-DMD approach can directly capture, from data, the breakdown of Floquet driven topological phases into Anderson localization.

**Interplay of Topology, Non-Hermitian Skin Effect, and Anderson Localization.** To investigate the interplay among topological boundary states, the non-Hermitian skin effect (NHSE) [51–53], and disorder-induced localization, we employ DMD to extract the spectral and spatial properties directly from the spatiotemporal data of the extended non-Hermitian SSH model (see the SM file). By introducing a Hatano-Nelson non-reciprocal pump ($\gamma$) and Anderson disorder ($W$) into the SSH model, we identify three distinct physical regimes under open boundary conditions (Fig. 4):

In **Phase A** (Robust Topological Phase, $\delta\kappa = 0.3, \gamma = 0.2, W = 0.5$), the system retains its topological robustness. By evaluating the IPR and the center of mass (COM) of the DMD-extracted modes, we find that the topological zero mode exhibits a markedly enhanced IPR, while the bulk modes remain extended with COMs distributed around the lattice center, albeit slightly shifted to the right owing to the NHSE. The spatial profiles further confirm edge localization of the left zero mode (Fig. 4d1). In this weak-disorder regime, the complex-energy spectrum and the reconstructed effective complex band structure remain close to the Bloch band picture (Fig. 4e1, f1), indicating that the effective wave vector is a well-defined quantum number.

In **Phase B** (NHSE-Dominant Phase, $\delta\kappa = 0.3, \gamma = 0.8, W = 0.5$), strong non-reciprocal pumping induces a macroscopic drift toward the right boundary. The COM distribution shows that nearly all DMD modes accumulate near the right edge (Fig. 4c2), while the isolated topological mode is no longer clearly separated in the IPR distribution of the extracted modes (Fig. 4b2). The spatial profiles of the DMD modes show pronounced accumulation near the right boundary, indicating the suppression of topological localization in the NHSE-dominant phase (Fig. 4d2). Strikingly, the DMD-extracted band structure strongly deviates from the periodic boundary condition (PBC) Bloch bands, providing direct, data-driven evidence of generalized Brillouin-zone deformation.

In **Phase C** (Anderson-Dominant Phase, $\delta\kappa = 0.3, \gamma = 0.2, W = 3.0$), strong disorder severely breaks translational symmetry, thereby suppressing both topological protection and the NHSE. The IPR of the DMD-extracted modes remains broadly enhanced (Fig. 4b3), but unlike the directional boundary accumulation seen in the NHSE phase, the COM of the DMD modes in this phase are randomly distributed throughout the bulk (Fig. 4c3), indicating disorder-induced localization rather than skin accumulation. Correspondingly, the DMD-extracted complex-energy spectrum becomes nearly real (Fig. 4e3), and the effective complex band structure loses the periodic coherence associated with the Bloch picture (Fig. 4f3). These results demonstrate that the Koopman-DMD framework can effectively disentangle the competing localization mechanisms associated with topological protection, the NHSE, and disorder scattering by data-tracking the dynamical changes in modal quantities such as COM and IPR.

**Data-Driven band topology in 2D Lattices.** To extend the Koopman-DMD framework from 1D chains to 2D lattices, we consider several representative 2D models, including the 2D SSH model as a prototypical higher-order topological insulator (HOTI) [54–56], graphene as a canonical Dirac system [36], the Haldane model as a time-reversal-symmetry-broken Chern insulator [35], and the Lieb and Kagome lattices as representative frustrated flat-band systems [57,58]. For these TBA models (see the SM file), the DMD extraction of band topology and quantum geometry—such as Berry curvature, Chern numbers, and quantum metrics— reveals subtle geometric structures that are otherwise hidden in standard macroscopic observables. In the following subsections, we show that DMD remains highly effective in capturing both real-space topological localization and momentum-space quantum geometry in the diverse 2D topological materials.

**Higher-Order Topological Phases.** The 2D SSH model generalizes the 1D dimerized chain to a dimerized square lattice and provides an intuitive realization of a higher-order topological insulator (HOTI), see the model description in the SM file. Although its first-order bulk topology, as characterized by Chern numbers and Berry curvature, is trivial, it supports nontrivial winding numbers that protect 1D edge states and, crucially, 0D corner states under open boundary conditions (OBC).

By applying DMD to spatiotemporal snapshots (Fig. 5a), we compare the extracted mode under open and periodic boundary conditions (OBC and PBC) (see Fig. 5b–f). Under OBC, some DMD modes lie outside the bulk band surfaces, and the energy–IPR distribution further shows a clear hierarchy, with corner, edge, and bulk states having the highest, intermediate, and lowest IPR values, respectively (Fig. 5b, c). Representative DMD mode profiles confirm that OBC supports bulk, edge, and corner modes, whereas under PBC only bulk modes remain on the band surfaces and no edge or corner modes are found (Fig. 5d–f). Furthermore, by analyzing the reconstructed bulk modes under PBC (Fig. 5m), DMD successfully extracts the nontrivial fractional winding numbers (quantized at $\pm\pi$) (Fig. 5n) and reproduces the trivial Berry curvature (Fig. 5r). The extracted quantum metric tensors ($g_{xx}, g_{yy}, g_{xy}$) also show excellent quantitative agreement with theoretical predictions (Fig. 5o–q). Collectively, these results reveal that our data-driven approach is highly sensitive to higher-order topological boundary conditions, capturing both the hierarchical localization of corner modes in real space and the subtle band geometric structure in momentum space without requiring exact diagonalization.

**From Gapless Graphene to the Topological Haldane Phase.** The graphene lattice forms a two-dimensional honeycomb structure whose low-energy excitations are described by the massless Dirac equation [36]. Although pristine graphene preserves both time-reversal and inversion symmetries and therefore remains globally topologically trivial, each Dirac valley carries a nontrivial Berry phase of π ($K$ and $K'$). Applying DMD to the spatiotemporal data under mixed boundary conditions, we successfully extract the characteristic linear Dirac-cone dispersion and the associated edge states (Fig. 6d–f, first panel). Furthermore, the DMD reconstruction captures the vanishing total Berry curvature, consistent with the globally trivial topology of pristine graphene, while the quantum metric shows singular enhancement near the Dirac points owing to the gapless band touching and the rapid variation of Bloch states in momentum space (Fig. 6a–f , second panel).

To realize a nontrivial topological phase, we consider the celebrated Haldane model [35], which introduces next-nearest-neighbor hopping that breaks time-reversal symmetry. This opens a topological gap governed by the massive Dirac equation and drives the system into a quantum anomalous Hall (QAH) insulating phase [59]. In this regime, the DMD analysis resolves the parabolic dispersion near the opened gap and captures the emergence of highly localized chiral edge states (Fig. 6a–f, third panel). The reconstructed quantum geometry remains strongly enhanced near the gapped Dirac points, reflecting the rapid variation of Bloch states inherited from the Dirac-type band structure, while the Berry-curvature integral over the Brillouin zone yields a quantized Chern number $C = 1$ (Fig. 6f, fourth panel). Together with the graphene case, these results show that our data-driven Koopman-DMD approach can directly distinguish a gapless Dirac semimetal from a topological Chern insulator. They further demonstrate the ability to reconstruct both the chiral boundary dynamics and hidden quantum geometry independently from finite-sample time-domain data, without requiring analytical Hamiltonian diagonalization. We also analyzed the flat-band topology in Lieb and Kagome lattices [60], which again shows excellent agreement with the exact results (see the SM file).

Taken together, the 1D and 2D lattice models investigated in this section provide a comprehensive and highly representative framework for exploring localization and topological band physics. Rather than forming a set of isolated examples, these TBA models are physically connected and complementary: the Anderson and SSH models establish the basic competition between disorder and topology, which is then extended to the non-equilibrium settings of Floquet driving and non-Hermitian physics. The extension to 2D systems further broadens the scope of our analysis by connecting the coexistence of bulk, edge, and corner modes in the 2D SSH model, the contrasting Dirac and Chern physics of graphene and the Haldane model, and the flat-band physics. By extracting energy spectra, spatiotemporal structures, and geometric tensors across these 1D and 2D lattice models, our results establish Koopman-DMD as a versatile data-driven framework for diagnosing localized and topological band physics directly from observational data.

We note that the benchmark models used here are not intended to demonstrate new phases in these models. Instead, they provide controlled ground truth for testing whether Koopman-DMD can recover spectral, localization, and quantum-geometric information from time-domain data. When a reliable Hamiltonian is known, conventional diagonalization remains the most direct route. The practical value of Koopman-DMD lies in data-first situations where the Hamiltonian is unknown, incomplete, model-dependent, or experimentally inaccessible.

**Koopman perspective on physical observables.**

Spatiotemporal data in quantum, optical, acoustic and nonequilibrium systems span several physical levels, ranging from state-resolved wavefunctions to strongly projected observable dynamics and collective responses. Accordingly, the appropriate variants of DMD [27] are determined primarily by the physical nature of the data rather than algorithmic preference, and the extracted DMD modes must be interpreted consistently with the underlying dynamical hierarchy. For state-level data, such as wavefunctions $\psi(x,t)$ and electromagnetic fields $E(x,t)$, the dynamics can often be described as an approximately closed evolution in Hilbert space. In this setting, exact DMD provides a natural data-driven approximation to the time-evolution operator (see the SM file), with eigenvalues that encodes band dispersion or Floquet quasienergy that can be related to single-particle states such as Bloch modes, Floquet harmonics.

We note that the benchmark models analyzed in the preceding sections are primarily single-particle lattice models. Their role is to validate the Koopman-DMD reconstruction of spectra, localization, and band geometry in controlled settings. Extending the framework to genuinely interacting many-body systems requires specifying what type of many-body spatiotemporal data are available. Depending on the data level, Koopman-DMD may be applied to time-dependent mean-field or effective fields, to finite many-body state vectors in Hilbert space, or to Green's functions. The physical interpretation of the extracted modes changes accordingly.

When only density and current fields, $\rho(x,t)$ and $j(x,t)$, are available, the dynamics has already been projected onto hydrodynamic observables and therefore carries only partial phase information. After a Madelung-type reconstruction of an effective wavefunction $\psi(x,t) =$ $\sqrt{\rho(x,t)}\exp\frac{i}{\hbar}\int\frac{mJ(x,t)}{\rho(x,t)}\cdot dx$ [61], exact DMD may still be applied, although time-delayed embedding (Hankel) DMD is more robust and physically appropriate because the projection induces effective memory. In this case, the extracted DMD modes are better interpreted as effective density waves, current waves, or scattering channels rather than as microscopic excitations.

For pump–probe spectroscopy data, such as transient optical, photoemission, or energy-loss spectra $S(\omega, t_{delay})$, the measured signal reflects the nonequilibrium response of driven systems and generally contains multiple coherence and relaxation timescales. Such data are intrinsically dissipative and non-Markovian, and cannot be described by a closed evolution in observable space. In this setting, Hankel DMD is particularly appropriate. It decomposes the pump–probe signal into modes with well-defined oscillation frequencies, damping rates, and spectral weights. These modes may then be interpreted as transient scattering channels, coherent relaxation pathways, or collective response components. For correlation functions, such as density–density and current–current correlations (see Table 1), the data represent time-domain response kernels associated with collective excitations. Their dynamics are fundamentally collective because they reflect a reduced description in terms of collective degrees of freedom. In this case, momentum-resolved Hankel DMD provides a data-driven reconstruction of the poles of the susceptibility $\chi(k,\omega)$. The DMD modes encode the dispersion and damping of collective modes, such as plasmons, spin waves, or order-parameter oscillations, while the projection weights represent their spectral residues. This provides one possible route toward analyzing many-body response data, because interacting systems are often more naturally characterized through Green's function than through single-particle wavefunctions.

Finally, for time traces of only a few of observables, such as $x(t), p(t)$, or spin components $S_\alpha(t)$, one is dealing with the most strongly projected form of quantum dynamics. These observables are low-dimensional expectation values of an underlying density matrix and therefore exhibit pronounced memory effects due to the elimination of hidden degrees of freedom. In this regime, Hankel DMD again provides an effective low-rank realization of the projected Koopman or Liouville dynamics. The extracted modes encode oscillation frequencies and decoherence or relaxation rates, while the corresponding mode components describe relative phase and amplitude structure across measurement channels.

Taken together, spatiotemporal datasets span multiple physical levels, from state-resolved wavefunctions to strongly projected observable dynamics. As the information content decreases and memory effects become more pronounced, the appropriate data decomposition tends to shift from exact DMD to more elaborate but expansive variants [27]. While DMD modes extracted from wavefunctions may be interpreted as quasiparticles, those obtained from observables, pump–probe spectra, correlations functions, and few-observable time traces are more appropriately understood as effective collective modes, scattering channels, or even poles of an effective susceptibility $\chi(k,\omega)$, rather than as direct representations of the single-particle spectral function $A(k,\omega)$.

**Final remarks**

In summary, Koopman–DMD provides a data-driven route to extracting effective modal excitations and spatiotemporal coherent structures from data without requiring a fully specified microscopic model. In the present work, this approach is validated on controlled single-particle lattice benchmarks and is particularly relevant to settings involving dissipation, disorder, nonequilibrium driving, or limited experimental access, where conventional Hamiltonian FBD approaches may be incomplete or impractical. At the same time, its capabilities remain constrained by data quality, finite spatiotemporal resolution and the fact that strongly interacting or strongly nonlinear dynamics may broaden spectra and obscure sharply defined DMD modes. For genuinely interacting many-body systems, the extracted DMD modes should be interpreted according to the type of input data, such as mean-field variables, many-body state vectors, or Green's functions, rather than automatically as single-particle band modes. The readout-noise benchmark in the SM provides an initial assessment of reconstruction robustness, while applications to specific experimental platforms require further evaluation of their measurement noise and uncertainties. Such platform-specific assessments, together with extensions to genuinely interacting many-body systems, remain important directions for future work.

From this perspective, Hamiltonian Floquet–Bloch decomposition and Koopman–DMD are best viewed as complementary rather than competing frameworks. Hamiltonian methods derive band structure, spectral functions, topological invariants, and quantum geometry from governing equations in a top-down manner, whereas Koopman–DMD infers effective frequencies, decay rates, and coherent mode profiles directly from observation in a bottom-up manner. Their complementarity is especially useful in driven, disordered, and partially unknown systems, where one framework supplies physical structure and the other supplies data-driven reconstruction. The present work validates this reconstruction strategy on benchmark models with known Hamiltonian reference solutions; applying it to genuinely unknown experimental system remains an important direction for future work. In this sense, Koopman–DMD provides a physically interpretable data-analysis framework that complements, rather than replaces, Hamiltonian band theory.

**Acknowledgments**

Y. P. acknowledges the support of the NSFC (No. 2023X0201-417-03) and the start-up fund from ShanghaiTech University.

Correspondence and requests for materials should be addressed to Y.P.: yiming.pan@shanghaitech.edu.cn.

**Data and code availability**

The code repository will be made publicly available on acceptance through an open-source link.

**Conflict of interest**

The authors declare no competing interests.

**Figures:**

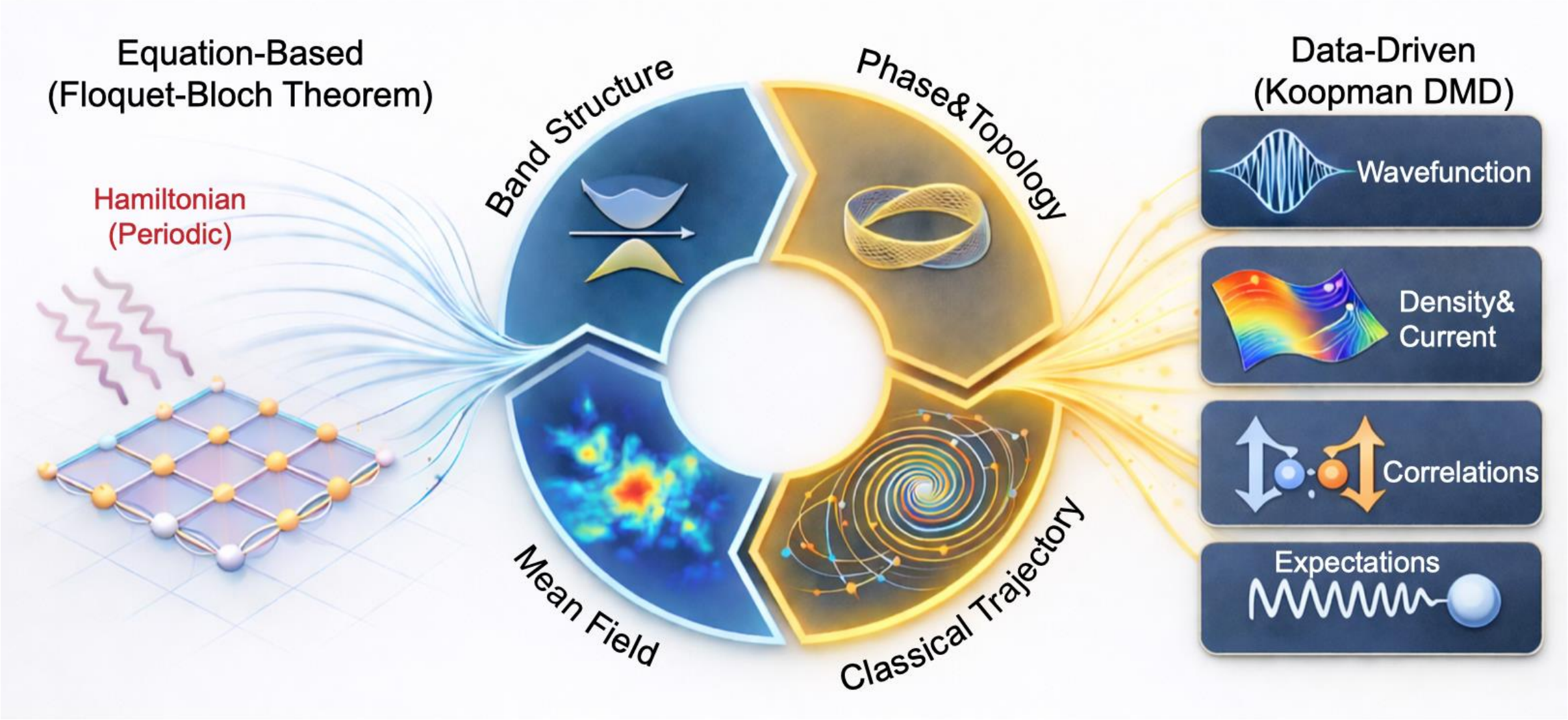

**Figure 1 | Complementary paradigms for analyzing spatiotemporal periodic systems.** A schematic comparison between the traditional equation-based framework and the data-driven approach. **Left:** The Hamiltonian-driven paradigm relies on the Floquet–Bloch theorem to derive band structures, topological invariants, and mean-field properties from an explicit time-periodic Hamiltonian. **Right:** The data-driven paradigm utilizes Koopman-operator theory and Dynamic Mode Decomposition (DMD) to infer Floquet-Bloch-like modes and band-related diagnostics from suitably resolved time-resolved observables (e.g., wavefunctions, densities, and correlations) without starting from an explicit microscopic model. Together, these two paradigms provide complementary routes between theoretical modeling and mode-resolved analysis of data.

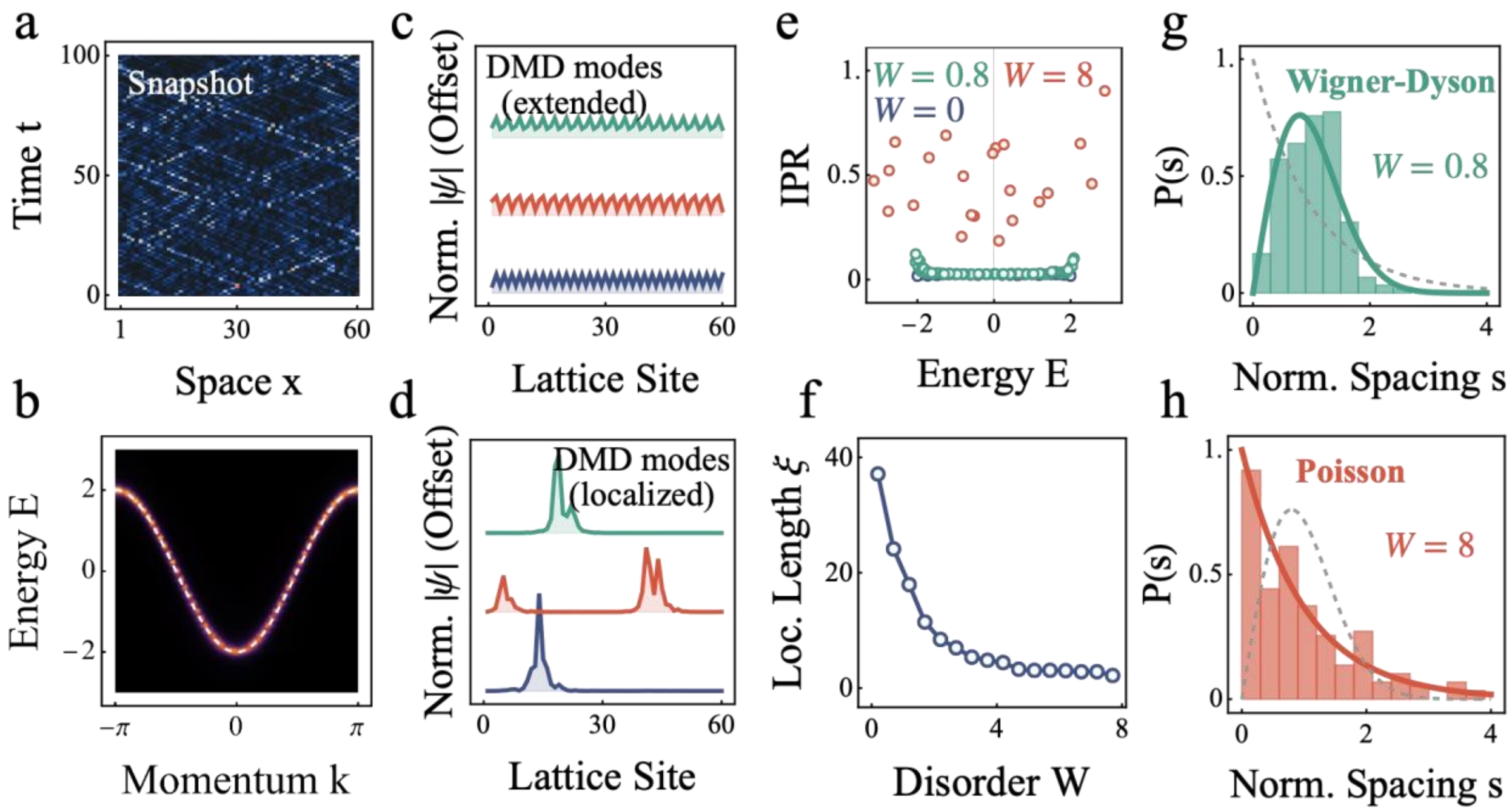

**Figure 2 | DMD analysis of the Anderson model and the crossover to localization.** (a) Spatiotemporal evolution data seeded from the random. (b) Band structure and spectral weights extracted by DMD. The dashed curves denote the exact dispersion. (c, d) Representative spatial profiles of the DMD modes in the weak- and strong-disorder regimes, showing extended and localized characteristics, respectively. (e) Distribution of the extracted modes in the energy–IPR plane under varying disorder strengths. (f) The DMD-extracted localization length decreases monotonically as the disorder strengthens. (g, h) Level-spacing statistics obtained from the DMD spectrum in the weak- and strong-disorder regimes, showing a transition from Wigner–Dyson to Poisson distributions.

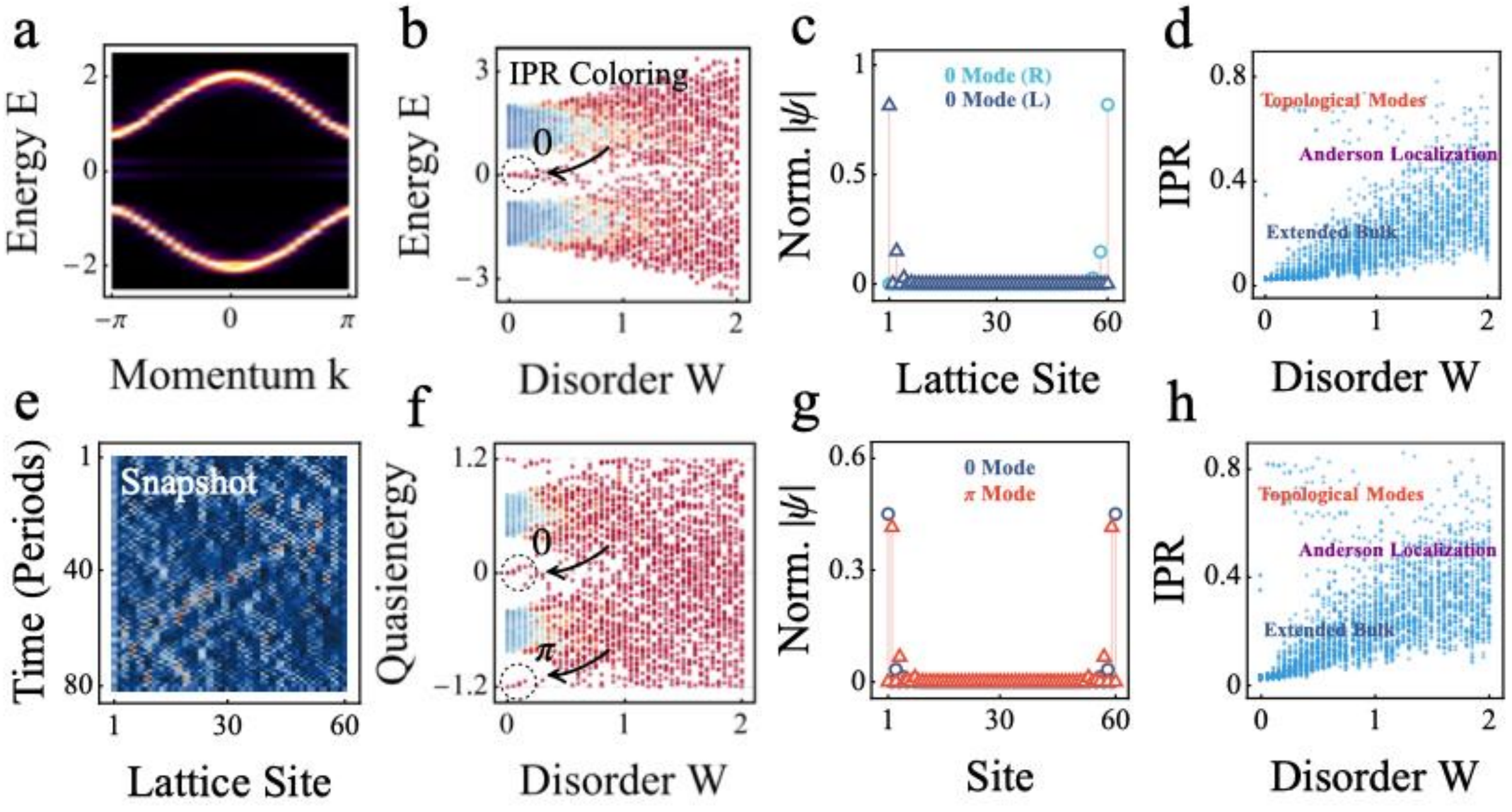

**Figure 3 | DMD analysis of the topological-to-Anderson localization transition in disordered static and Floquet SSH models.** Panels **(a)–(d)** correspond to the static disordered SSH model, while **(e)–(h)** present the Floquet SSH model. **(a)** The spectral function extracted by DMD. **(b)** The energy spectrum with disorder strength $W$, colored with the IPR. Both the topological and bulk modes are gradually absorbed into the Anderson regime. **(c)** Spatial profiles of the extracted edge-localized zero modes. **(d)** Inverse Participation ratio (IPR) versus $W$, signaling topological-to-Anderson transition. **(e)** Spatiotemporal input data for the Floquet SSH model. **(f)** The energy spectrum with disorder strength $W$, colored with the IPR, revealing both 0- and $\pi$-modes are absorbed into the Anderson regime. **(g)** Spatial profiles of the 0- and $\pi$-modes. **(h)** The IPR versus $W$, demonstrating the transition from Floquet topological phase to Anderson regime.

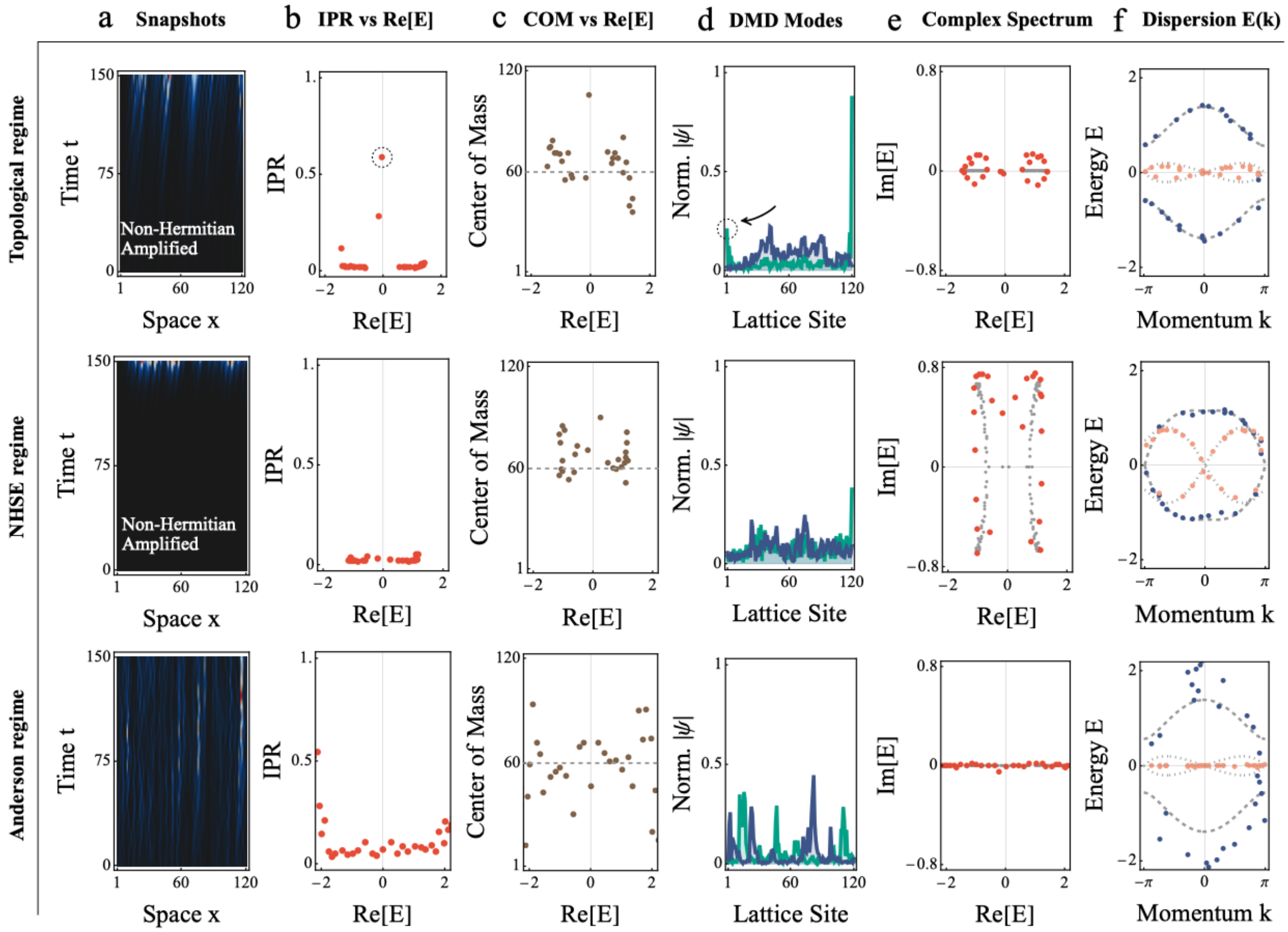

**Figure 4 | DMD analysis of topological and localization transitions in the disordered non-Hermitian SSH model.** Each column corresponds to a specific analysis, labeled **(a)–(f)** from left to right. The three rows represent Phase A (topological), Phase B (NHSE-dominant), and Phase C (Anderson-dominant), respectively. **(a)** Spatiotemporal amplitude evolution under wideband random excitation. **(b, c)** IPR and Center of Mass (COM) of the DMD-extracted modes as a function of $\mathrm{Re}(E)$. The COM distribution clearly distinguishes topological edge pinning, NHSE-induced boundary accumulation, and random Anderson localization. **(d)** Spatial profiles of representative DMD modes, highlighting the competition between the topological zero mode (green) and bulk modes (blue) under non-Hermitian asymmetry. **(e)** Complex-energy spectrum under open boundary conditions, comparing exact diagonalization (gray) with DMD-extracted modes (red). **(f)** Effective complex band structure reconstructed via DMD, highlighting the strong deviation from theoretical Bloch bands caused by GBZ deformation and disorder scattering.

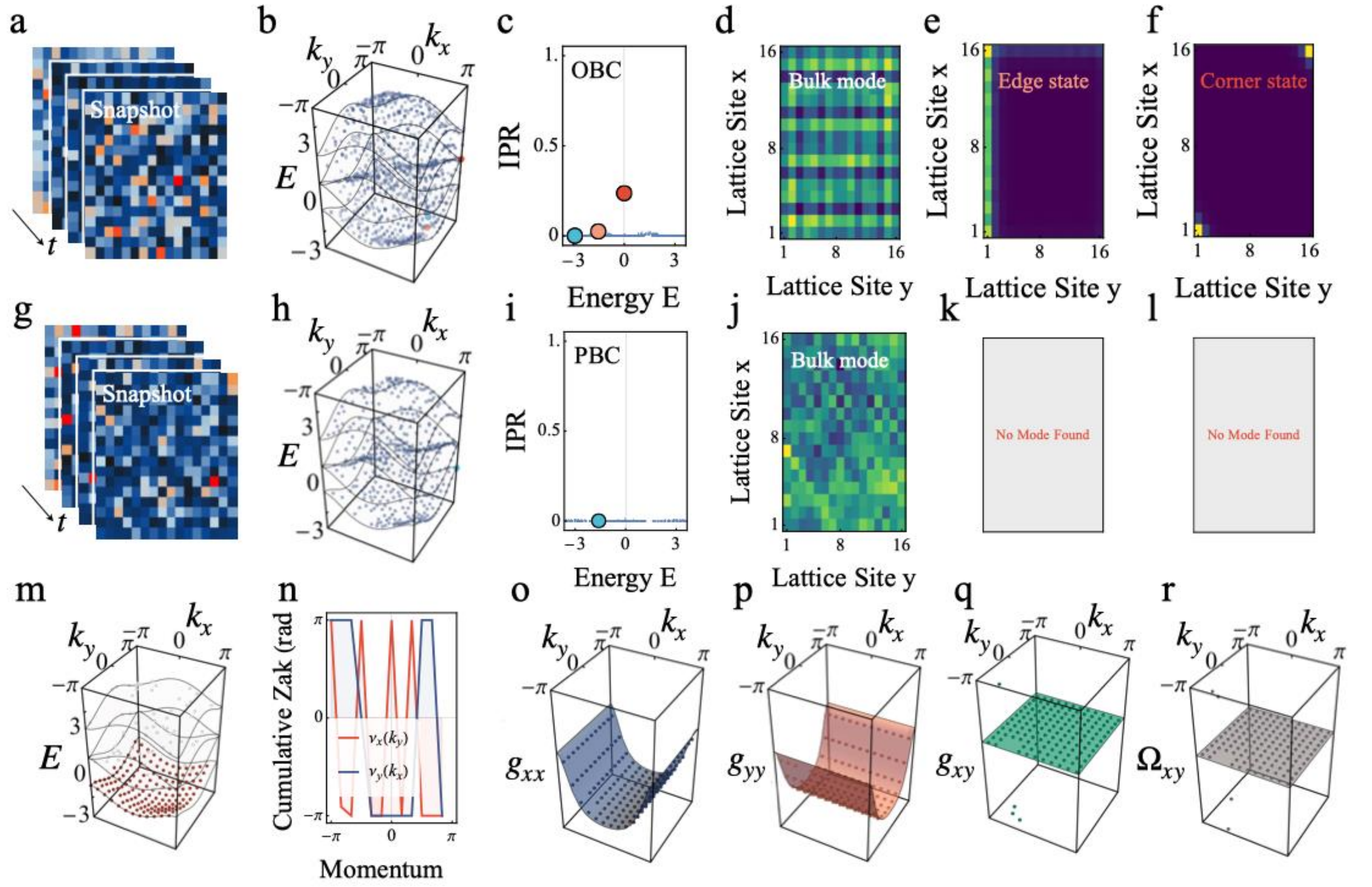

**Figure 5 | DMD analysis of the higher-order topological phases and quantum geometry. (a)–(l)** compare the results under open boundary conditions (upper row) and periodic boundary conditions (middle row), and **(m)–(r)** present the topological analysis of the extracted modes. **(a, g)** Spatiotemporal snapshots used as input for DMD. **(b, h)** DMD-extracted modes overlaid on the theoretical band surfaces, distinguishing bulk, edge, and corner states under OBC. **(c, i)** Energy–IPR distribution of the extracted modes, quantifying the spatial confinement hierarchy. **(d)–(f), (j)–(l)** Spatial profiles of representative bulk, edge, and corner modes. **(m)** Bloch modes selected from a full band for topological analysis. **(n)** Winding numbers obtained from DMD, quantized at $\pm\pi$ in the nontrivial phase. **(o)–(q)** Comparison between the DMD-extracted quantum metric components (black dots) and theoretical surfaces, demonstrating high-fidelity geometric reconstruction. **(r)** Reconstructed Berry curvature, confirming a vanishing total sum.

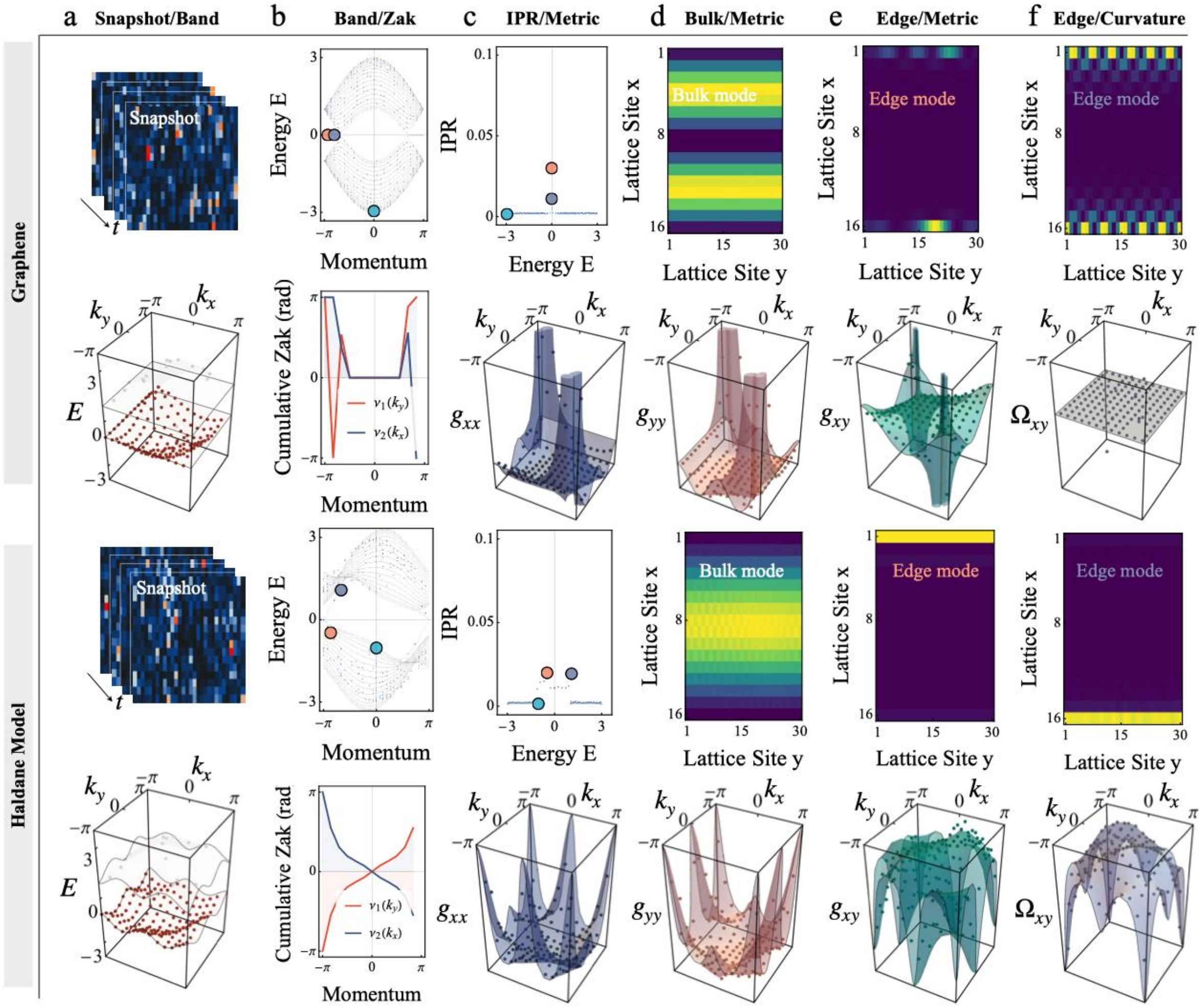

**Figure 6 | Comparative DMD analysis of topological phases and quantum geometry in graphene and Haldane models.** The figure is divided into two main panels indicating the models: pristine Graphene (top rows) and the Haldane model (bottom rows). Columns **(a)–(f)** display the corresponding data-driven mode extraction and topological analysis for both models. **(a)** *Snapshot/Band*: (Upper) Spatiotemporal snapshots used as input for DMD. (Lower) Reconstructed 3D energy band structures. **(b)** Band/Zak: (Upper) DMD-extracted modes overlaid on theoretical band structures. Graphene exhibits a gapless Dirac cone, whereas the Haldane model shows an opened topological gap with chiral edge states (red markers). (Lower) Calculated cumulative Zak phases. **(c)** IPR/Metric: (Upper) Energy–IPR (Inverse Participation Ratio) distributions distinguishing bulk states from highly localized edge states. (Lower) Reconstructed quantum metric component $g_{xx}$. **(d)–(e)** Bulk & Edge/Metric: Spatial profiles of representative bulk (d, upper) and edge (e, upper) modes, alongside their corresponding reconstructed quantum metric components $g_{yy}$ (d, lower) and $g_{xy}$ (e, lower). **(f)** Edge/Curvature: (Upper) Spatial profile of edge modes. (Lower) Reconstructed Berry curvature $\Omega_{xy}$. The results confirm the globally trivial topology (zero net curvature) of graphene, while the Brillouin-zone integral of the Berry curvature for the Haldane model yields $C = 1$, characteristic of the QAHE phase.

| Physical Observables | Physical Information (Prior) | DMD Space | DMD Method | Physical Information from Data (Posterior) |
|---|---|---|---|---|
| $\psi(x,t)$, or $\boldsymbol{E}(x,\ t)$ | quantum or classical state dynamics | state space | Exact | Floquet-Bloch modes and band structure |
| $\rho(x,t),\ \mathbf{j}(x,t)$ | dynamical density distribution | observable space | Exact/Hankel | single particle transport |
| $I(\omega,\tau)$ | pump-probe and nonequilibrium response | observable space | Hankel | scattering matrix element |
| $\langle jj\rangle$, $\langle S_\alpha S_\beta\rangle$ | correlation and collective response | observable space | Hankel | collective excitation and order parameter |
| $\langle x\rangle$, $\langle p\rangle$, $\langle S_\alpha\rangle$ | semiclassical motion of expectation value | state/observable space | Exact/Hankel | classical trajectory and phase portrait |

**Table 1 | Classification of physical observables and extraction methods in the Koopman-DMD framework.** This table categorizes various types of spatiotemporal data—ranging from fundamental wavefunctions to projected physical observables—and outlines their corresponding DMD reconstruction strategies. For full state dynamics (e.g., wavefunctions ψ), exact DMD extracts spatial eigenmodes such as Floquet-Bloch modes. In contrast, for strongly projected observables (e.g., density/current correlations, pump-probe responses, and semiclassical expectation values), the dynamics are generically non-Markovian. In these regimes, time-delayed (Hankel) DMD provides a robust data-driven realization of the extended Koopman operator. The extracted eigenvalues encode oscillation frequencies and decay rates, while the eigenmodes capture spatiotemporal coherent structures. Ultimately, since Koopman operator governs the linear evolution of observables, this framework provides a natural and powerful tool for reconstructing underlying physics directly from diverse experimental measurements.

## Supplementary Material:

# Data-driven reconstruction of band dispersion and quantum geometry via Koopman dynamic mode decomposition

Yiming Pan[1*], Jinze He[1], Jiapeng Yang[1], Zhiwei Fan[2]

*[1]State Key Laboratory of Quantum Functional Materials, School of Physical Science and Technology and Center for Transformative Science, ShanghaiTech University, Shanghai 200031, China*

*[2]School of Engineering, Newcastle University, Newcastle upon Tyne, NE1 7RU, UK*

This SM file includes five sections:

- Section S1: Koopman DMD framework and workflow
- Section S2: Tight-Binding Models, Band Analysis, and Topological Phase Transitions from Hamiltonian FBD
- Section S3: Readout noise and reconstruction robustness
- Section S4: Quasiparticle spectral function extracted from DMD.
- Section S5: Koopman perspective on Green's functions and propagators

**Section S1: Koopman DMD framework and workflow**

To analyze band-related structure in complex spatiotemporal data, we employ the Koopman dynamic mode decomposition (DMD) framework. Rooted in Koopman operator theory, DMD provides a data-driven finite-dimensional approximation to the observed evolution of sampled spatiotemporal data. By decomposing the data into coherent modes, DMD yields modal frequencies, effective decay rates, spatial profiles, and projection weights. For sufficiently resolved state-level data, these quantities can be used to infer band dispersion, localization diagnostics, and selected geometric quantities.

**A. Spatiotemporal Data Acquisition and Preprocessing**

Before employing DMD, the physical fidelity of the input data should be assessed. The reliability of the extracted modal frequencies, spatial structures, and momentum-resolved diagnostics is limited by the temporal sampling interval, spatial sampling interval, total observation time, and system size, excitation coverage, and the selected truncation rank. According to the Nyquist-Shannon sampling theorem, the discrete temporal sampling interval $\Delta t$ and spatial lattice spacing $\Delta x$ fundamentally define the maximum observable frequency $\omega_{max} = \pi/\Delta t$ and momentum $k_{max} = \pi/\Delta x$, respectively. The total observation time $T$ and the spatial system size $L$ dictate the minimal resolution limits: $\delta\omega = 2\pi/T$ and $\delta k = 2\pi/L$. These quantities define the accessible frequency-momentum window and its finite resolution. DMD modes whose frequencies or wavevectors fall outside these sampling bounds, or whose apparent fine structure lies below the corresponding resolution limits, are treated as unreliable for quantitative interpretation and are excluded from the final reconstruction.

Experimental measurements and numerical simulations may also contain systematic biases, static backgrounds, or ultra-low-frequency temporal drifts unrelated to the target dynamic modes of interest. When such components are present, they may be removed by standard detrending or background-subtraction procedures such as utilizing Principal Component Analysis (PCA) and Empirical Mode Decomposition (EMD) [S1,S2]. Unlike Koopman-DMD, PCA and EMD are used here only as preprocessing or empirical component-separation tools; they do not by themselves yield Koopman eigenvalues, modal decay/growth rates, projection weights, or a direct Floquet-Bloch-like modal interpretation. Such preprocessing should be used conservatively and validated against the unprocessed data whenever geometric or topological quantities are computed.

**Noise and Robustness Analysis.** Even for noise-free input data, DMD may exhibit numerical artifacts arising from finite temporal sampling, a limited observation window, and SVD rank truncation, rather than from the SVD itself. Insufficient temporal sampling may cause aliasing, whereas a short observation window limits frequency resolution and can lead to mode mixing. We therefore adopt three complementary strategies: (i) choosing the sampling interval and observation duration to resolve the relevant dynamical scales; (ii) applying an appropriate singular-value threshold and retaining a sufficiently low rank to preserve robust physical modes while suppressing weak, noise-dominated components; and (iii) repeating the analysis for different initial states or datasets to identify reproducible modes and improve modal coverage. These procedures reduce, but do not eliminate, artifacts associated with finite sampling and rank truncation.

We address that disorder constitutes a physical perturbation distinct from numerical or measurement noise and provides a direct test of the robustness of the DMD reconstruction. For the disordered SSH model shown in Fig. 3, the topological modes remain resolvable at weak disorder. As the disorder strength increases, these modes progressively hybridize with disorder-induced background states and lose their spectral and spatial contrast. Thus, the relevant physical features can still be recovered when disorder-induced mode mixing remains limited. Once the modes are no longer distinguishable from the background, however, their identification by DMD becomes unreliable.

Experimental noise differs from the effects above because it directly perturbs the measured spatiotemporal snapshots. Weak, predominantly low-amplitude noise may be suppressed by through appropriate SVD rank truncation, provided that the extracted modes remain stable against reasonable variations of the truncation threshold. When the noise becomes comparable to or larger than the physical signal, however, rank truncation alone cannot recover the lost information. The experimental signal-to-noise ratio must then be improved, for example through higher measurement precision or repeated acquisition, before DMD can reliably resolve the intrinsic dynamic modes and coherent spatiotemporal structures.

**Excitation Strategies for data preparation**: A single, generic spatiotemporal dataset is generally insufficient to reveal all dynamically accessible sectors of the Brillouin Zone (BZ). To comprehensively collect the band structure across all momentum and energy sectors, we strategically design specific excitation profiles to probe targeted regimes:

(i) Single-point Excitation:

Acting as a spatial delta function, this strategy provides a broad, flat spectrum in momentum space, efficiently injecting energy into all $k$-states simultaneously to yield a rapid overview of the available bands.

(ii) Random/White-noise Excitation:

By introducing uncorrelated spatiotemporal perturbations, this approach provides an unbiased, homogenous energy injection across both frequency and spatial domains. It is highly optimal for mapping the complete continuum of bulk energy bands without favoring specific symmetries.

(iii) Targeted Bulk vs. Edge Excitation:

To distinguish between trivial and non-trivial states, the spatial location of the excitation is carefully controlled. Bulk excitations are deliberately applied deep within the lattice to extract the interior continuous dispersion while circumventing boundary effects. By contrast, localized edge excitations are engineered to couple exclusively with the boundaries and interfaces, allowing for the high-contrast isolation and illumination of topologically protected edge states without interference from the bulk continuum.

**Validation and convergence with data size.** In the present benchmarks, the DMD reconstruction is validated by comparing the extracted frequencies, mode profiles, IPR and COM diagnostics, topological invariants, and quantum-geometric quantities with independent Hamiltonian-based calculations. Because DMD is used here as a deterministic modal decomposition of the available snapshots rather than as a supervised predictive model, validation is based on agreement with these reference calculations rather than on a separate training-validation split. Data sufficiency and numerical convergence are assessed by varying the total observation time, temporal sampling interval, system size, excitation coverage, and SVD truncation rank. For experimental or otherwise unknown systems without an explicit reference model, robustness of the extracted modes may instead be evaluated using repeated measurements, independent datasets, and their stability against reasonable variations in the control parameters.

### B. Exact DMD and Hankel DMD Framework

Let the spatiotemporal evolution of the system be represented by a snapshot matrix $X = [x_1, x_2, \dots, x_{N-1}]$, where $x_i$ represents the system state at time $t_i$, and $Y = [x_2, x_3, \dots, x_N]$ represents the time-shifted state.

**Exact DMD**: The standard Exact DMD seeks the best-fit linear operator $A$ such that $Y \approx A\,X$. To compute this efficiently, we perform a Singular Value Decomposition (SVD) on the data matrix, $X \approx U\Sigma V^{\dagger}$, truncated to a rank $r$. The operator $A$ is then projected onto the reduced subspace spanned by $U$, giving $\tilde{A} = U^{\dagger} Y V \Sigma^{-1}$. Solving the reduced eigenvalue problem, $\tilde{A} w_j = \lambda_j w_j$, yields the DMD eigenvalues $\lambda_j$, from which the Exact DMD modes in the original space are reconstructed as $\Phi_j = Y V \Sigma^{-1} w_j$.

**Hankel DMD** (time-delay embedding): In some systems, a single snapshot may not provide a sufficiently informative representation of the effective dynamics, particularly when hidden variables, limited observables, or finite-time spectral mixing are important. In such cases, we use Hankel DMD, in which $q$ consecutive snapshots are stacked to form a delay-embedded matrix,

$$H = \begin{bmatrix} \mathrm{x}_0 & \mathrm{x}_1 & \cdots & \mathrm{x}_{m-1} \\ \mathrm{x}_1 & \mathrm{x}_2 & \cdots & \mathrm{x}_m \\ \vdots & \vdots & \ddots & \vdots \\ \mathrm{x}_{q-1} & \mathrm{x}_q & \cdots & \mathrm{x}_{q+m-1} \end{bmatrix} \tag{S1}$$

Applying the DMD procedure to the shifted Hankel matrices $H_0$ and $H_1$ , the algorithm captures the phase-space geometry and extracts dressed quasiparticle excitations that inherently include correlation and memory effects.

The choice of Hankel-DMD parameters affects the quality of the reconstruction. For band reconstruction and spectral mapping, one typically chooses the embedding length so that $q\Delta t$ covers at least several oscillation periods of the dominant modes. In practice, one may begin with $q$ corresponding to roughly 5-20-time samples and increase it near low-frequency or gap-edge structures if necessary. The model rank $r$ should remain modest; for two-band systems such as SSH, a small rank is usually sufficient, while excessively large $r$ values tend to fit noise and generate artificial peaks. Mild temporal windowing of $\psi(k,t)$ may also reduce boundary effects, although it should be applied consistently if spectral weights are to be compared quantitatively.

### C. Extraction of Spectral Properties

The outputs of DMD are naturally connected to the spectral content of the measured dynamics. For each DMD mode $j$, the discrete eigenvalue $\lambda_j$ is mapped to the continuous-time domain via $s_j = \ln(\lambda_j)/\Delta t$. The complex frequency is expressed as $s_j = -\gamma_j - i\omega_j$. Here, $\omega_j = -\mathrm{Im}(s_j)$ denotes the oscillation frequency (energy) of the mode, and $\gamma_j = -\mathrm{Re}(s_j)$ is the decay rate corresponding to the imaginary part of the self-energy induced by inelastic scattering or radiation. The DMD Modes ( $\Phi_j$ ) encode the associated spatial structures of the extracted dynamical components. Their projection weights are obtained from the initial condition through the pseudoinverse relation $\boldsymbol{w} = \Phi^{+}x_1$.

### D. Reconstruction of physical information

The spectral decomposition $X(t) \approx \sum_{j=1}^{r} w_j \Phi_j\, e^{(-\gamma_j - i\omega_j)t}$ provides a compact representation from which several physically relevant diagnostics can be estimated.

**Band Structure and Q Factor:** The band structure (dispersion relation) is reconstructed by performing a spatial Fourier transform on the DMD modes $\Phi_j(r)$ to identify their dominant Bloch wavevector $k_j$. By plotting the extracted mode frequency $\omega_j$ against $k_j$, the data-driven energy bands $E(k)$ are formed. Furthermore, the Quality factor (Q factor) of each mode in the band is directly evaluated as $Q_j = |\omega_j|/(2\gamma_j)$, quantifying the resonance sharpness.

**Inverse Participation Ratio (IPR):** To characterize the spatial localization of the eigenstates—critical for identifying Anderson or many-body localization regimes—we compute the IPR for each normalized DMD mode:

$$\mathrm{IPR}_j = \frac{\sum_i |\Phi_j(r_i)|^4}{\left(\sum_i |\Phi_j(r_i)|^2\right)^2} \tag{S2}$$

An IPR approaching 1 indicates a highly localized state, while an IPR scaling as $1/L$ (where $L$ is the system volume) indicates an extended state.

**Spectral Function and Local Density of States (LDOS):** The discrete poles extracted by DMD allow for an analytical, noise-free reconstruction of the dynamical spectral function $A(\omega)$, circumventing the Gibbs ringing typical of discrete Fourier transforms on finite data:

$$A(\omega) = \sum_{j=1}^{r} |w_j|^2 \frac{\gamma_j}{(\omega-\omega_j)^2+\gamma_j^2} \tag{S3}$$

The LDOS at a specific spatial location $r$ is further obtained by spatially weighting this distribution: $\rho(r,\omega) = \sum_j |\Phi_j(r)|^2 \frac{\gamma_j}{(\omega-\omega_j)^2+\gamma_j^2}$.

To ensure the results remain physically meaningful, one must be cautious of the following limitations: (i) If the input is a generic complex amplitude, we may reconstruct frequencies but not necessarily the retarded spectral function unless you enforce causality (e.g., using only $(t > 0)$, appropriate windowing). (ii) A finite time window limits frequency resolution. DMD may sharpen modal fitting, but it cannot overcome the information bound set by the data duration. Noise and mode mixing may require more robust variants such as total-least-squares DMD, optimized DMD, or Hankel DMD. (iv) In interacting systems, a DMD analysis of a mean-field order parameter or reduced field generally yields a useful dynamical spectrum, but not in general the single-particle spectral function. (v) For Bloch-like modes, consistent phase and gauge tracking across momentum space is required when interpreting mode weights, Berry phases, or related geometric quantities as discussed below.

**Topological Invariants via Fukui-Hatsugai-Suzuki (FHS) Method:** The topological properties of the reconstructed band structure can be determined directly from the data-driven eigenstates extracted by the DMD algorithm.

By performing a spatial Fourier transform on the spatial profiles of the DMD modes $\Phi_j(r)$, we map them into momentum space to obtain the discrete Bloch wavefunctions $|u_n(k)\rangle$ for the $n$-th energy band. To compute the topological invariants while strictly maintaining gauge invariance on a discretized Brillouin zone grid, we employ the Fukui-Hatsugai-Suzuki (FHS) algorithm.

For one-dimensional (1D) systems, the topological characteristics of an isolated band are captured by the Zak phase (1D Berry phase) [S3], computed as:

$$\gamma = -\mathrm{Im}\ln\left(\prod_j \langle u(k_j) | u(k_{j+1}) \rangle\right) \tag{S4}$$

where the product runs over a closed discrete loop across the 1D Brillouin zone.

For two-dimensional (2D) systems, the Chern number $C$ is calculated by summing the Berry flux $F_{xy}$ over all elementary plaquettes in the discretized 2D momentum lattice. The Berry flux through a specific plaquette located at $k$ is given by the principal value of the phase of the product of overlap integrals around its boundary:

$$C = \frac{1}{2\pi} \sum_{\text{plaquettes}} \operatorname{Im} \ln\left(\langle u_k | u_{k+\delta x} \rangle \langle u_{k+\delta x} | u_{k+\delta x+\delta y} \rangle \langle u_{k+\delta x+\delta y} | u_{k+\delta y} \rangle \langle u_{k+\delta y} | u_k \rangle\right) \quad \text{(S5)}$$

This data-driven application of the FHS framework provides a robust and gauge-invariant topological classification directly from the system's empirical spatiotemporal evolution, seamlessly translating DMD eigenvectors into condensed matter topological invariants.

Notably, there is an alternative method to extract the topological phase. An alternative viewpoint is that the effective Koopman operator defines a non-Hermitian reduced operator $H_{\text{eff}} = i\ln(\tilde{A})/\Delta t$ based on the low-dimensional matrix $\tilde{A}$. Because $H_{\text{eff}}$ is non-Hermitian, it yields a biorthogonal basis consisting of right eigenvectors $|\boldsymbol{R_n}\rangle$ and left eigenvectors $\langle L_n|$. By tracking the evolution of these biorthogonal eigenstates across the reconstructed momentum space (parameter space), we compute the non-Hermitian Berry connection $A_n = -i\langle \boldsymbol{L_n} | \nabla_k | \boldsymbol{R_n} \rangle$. Integrating this connection over the Brillouin zone yields the Zak phase (for 1D topological lattice models) or the Chern number, enabling a data-driven classification of the system's topological phase.

**Quantum geometric tensor via DMD modes:** Beyond the macroscopic topological invariants, the local geometric structure of the reconstructed band is encoded in the quantum geometric tensor (QGT) $T_{\mu\nu}^{n} = \langle \partial_\mu u_n | (1 - |u_n\rangle\langle u_n|) | \partial_\nu u_n \rangle$, where n is the band index. The imaginary part of the QGT gives the **Berry curvature**, while its real part gives the **quantum metric,** which characterizes the distance between quantum states in Hilbert space. To extract these local geometric quantities from the gauge-dependent DMD mode $|u_k\rangle$ while maintaining strict gauge invariance, we extend the discrete FHS formalism. For a 2D discretized momentum grid with spacings $\delta k_x$ and $\delta k_y$, we define the elementary overlap integrals (link variables) along the principal axes as $U_x(k) = \langle u_k | u_{k+\delta x} \rangle$ and $U_y(k) = \langle u_k | u_{k+\delta y} \rangle$, alongside a diagonal cross-overlap $U_{xy}(k) = \langle u_{k+\delta x} | u_{k+\delta y} \rangle$. The local Berry curvature $\Omega_{xy}(k)$ is derived from the geometric phase accumulated around an elementary plaquette. Using the link variables, the gauge-invariant Berry flux through a specific plaquette located at $\boldsymbol{k}$ is given by the principal value of the phase:

$$\Omega_{xy}(k) = \operatorname{Im} \ln\left[U_x(k) U_y(k + \delta k_x) U_x^*(k + \delta k_y) U_y^*(k)\right] \quad \text{(S6)}$$

The Berry curvature is then strictly defined as $\Omega_{xy}(k) = F_{xy}(k)/(\delta k_x \delta k_y)$. Simultaneously, the diagonal components of the quantum metric tensor $g_{\mu\nu}$, representing the state distance along a single momentum direction, are calculated directly from the norm of the link variables:

$$g_{xx}(k) = \frac{1-|\langle u_k|u_{k+\delta x}\rangle|^2}{\delta k_x^2} \tag{S7}$$

$$g_{yy}(k) = \frac{1-|U_y(k)|^2}{\delta k_y^2} \tag{S8}$$

The off-diagonal components, which capture the non-orthogonal shear deformation between different momentum directions, are symmetric ( $g_{xy} = g_{yx}$ ) and are evaluated using the cross-overlap to explicitly compute the geometric projection:

$$g_{xy}(k) = \frac{\mathrm{Re}[U_{xy}(k)-U_x^*(k)U_y(k)]}{\delta k_x \delta k_y} \tag{S9}$$

This fully discretized formulation ensures that any arbitrary and disordered $U(1)$ phase factors—inherent to the independent eigenvalue decompositions performed by the DMD algorithm at each momentum point—perfectly cancel out. Consequently, it translates the empirical spatiotemporal evolution matrices into smooth, physically meaningful landscapes of the Berry curvature and a positive semi-definite quantum metric tensor, allowing gauge-invariant identification of topological phases and phase transitions without starting from an explicit Hamiltonian model.

**Momentum-Space DMD for Data Extraction**: To evaluate the topological invariants using the FHS algorithm, it is imperative to extract the cell-periodic Bloch functions $|u_n(k)\rangle$ with well-defined phase relationships across the Brillouin zone. Rather than performing DMD on the entire spatial lattice which may yield standing waves due to finite-size effects, we project the spatiotemporal data into momentum space prior to modal decomposition.

For a lattice with $N$ unit cells and α sublattices (orbitals) per cell, the state at unit cell $R$ and time $t$ is denoted as $x_\alpha(R,t)$. We first perform a spatial discrete Fourier transform to decouple the momentum sectors: $x_\alpha(k,t) = \frac{1}{\sqrt{N}}\sum_R e^{-ik\cdot R} x_\alpha(R,t)$. For each discrete momentum $k$ in the Brillouin zone, this yields a reduced time series vector $x(k,t)$ of dimension equal to the number of sublattices. By applying the Hankel DMD algorithm independently to each $x(k,t)$ time series, we extract the eigenvalues representing the band dispersion $E_n(k) = \mathrm{Re}(\omega_n(k))$, and the corresponding eigenvectors constitute the Bloch states $|u_n(k)\rangle$. These states are then sorted by energy to form continuous bands, directly serving as the input for the FHS gauge-invariant integrations.

### E. Optimal Rank Truncation in DMD:

The selection of the truncation rank $r$ during the Singular Value Decomposition (SVD) step matters in distinguishing intrinsic physical bands from numerical noise. Retaining an excessive number of singular values introduces spurious, unphysical modes (overfitting), whereas over-truncating discards essential macroscopic band features. To select $r$ in a reproducible manner, we employ a combined criterion of cumulative singular value energy (e.g., retaining $> 99\%$ of the total variance) and the optimal hard threshold (OHT) algorithm. This choice provides a compact reduced representation that preserves the dominant coherent structures in the data while suppressing weak components that are unlikely to be robust under finite-data conditions.

### F. Computational complexity and memory scaling

For a standard DMD analysis, let the snapshot matrix be $X \in \mathbb{C}^{N\times M}$ ,where $N$ is the number of spatial degrees of freedom, including sites, orbitals, sublattices, pixels, or field components, and $M$ is the number of temporal snapshots. The dominant computational cost is the singular-value decomposition of $X$. A dense full SVD scales as $O(NM^2)$ $(N \geq M)$, or $O(N^2M)$ $(M \geq N)$. In practice, DMD is performed with a rank-$r$ truncated SVD, $r \ll min(N, M)$, for which the leading cost is approximately $O(NMr)$. After the projection, the fitted Koopman/DMD operator is reduced to an $r \times r$ matrix, whose eigen decomposition costs $O(r^3)$. The leading memory requirement of standard DMD is $O(NM)$, corresponding to storage of the snapshot matrix, plus lower-order storage for the reduced matrices.

For Hankel DMD with delay depth $d$, the delay-embedded snapshot matrix has an effective row dimension of approximately $dN$ and an effective number of columns of order $M - d + 1$. Thus, the memory requirement scales approximately as $O\big(dN(M - d + 1)\big)$, and the leading truncated-SVD cost scales approximately as $O(dN(M - d + 1)r)$. This additional cost is the price paid for resolving memory effects, hidden variables, or strongly projected observables. Therefore, the choice of $r$ and $d$ must balance spectral accuracy, noise suppression, and computational feasibility.

These estimates clarify that Koopman-DMD is not computationally cost-free. Its practical advantage is that it avoids constructing and diagonalizing a full Hamiltonian when only data are available, but it still requires storage and factorization of the snapshot matrix. Low-rank or randomized SVD is therefore essential for large spatiotemporal datasets.

### G. Comparative workflow between Hamiltonian FBD and Koopman DMD.

Generally, the conventional Hamiltonian-based topological analysis proceeds as follows:

- **Model Construction:** Define the system's static Hamiltonian $H(\boldsymbol{k})$, or the time-periodic Hamiltonian $H(t)$ transformed into the Floquet-Bloch form.
- **Band Structure Calculation:** Solve the eigenvalue problem $H(k)|u_{nk}\rangle = E_n(k)|u_{nk}\rangle$ to obtain the energy dispersion $E_n(\boldsymbol{k})$. For periodically driven systems, diagonalize the one-period Floquet evolution operator $U(T)$ to extract the Floquet quasi-energies and modes.
- **Topological Invariants:** Compute topological quantities utilizing the obtained Bloch eigenstates, such as integrating the Berry connection to yield the Zak phase $\gamma_n = \int_{\mathrm{BZ}}\langle u_{n\boldsymbol{k}}|i\partial_{\boldsymbol{k}}|u_{n\boldsymbol{k}}\rangle d\boldsymbol{k}$, or the Berry curvature $\Omega_n(\boldsymbol{k})$ for the Chern number.

- **Spectral Functions:** Construct the retarded Green's function and the corresponding spectral function $A(\boldsymbol{k}, \omega)$ directly from the band dispersion and eigenstates to analyze quasiparticle excitations.

By comparison, the Koopman-DMD workflow proceeds in the opposite direction:

- **Data Acquisition:** Collect spatiotemporal snapshots of wavefunctions, fields, densities, currents, correlation functions, or expectation values.
- **Effective Evolution Operator:** Construct the snapshot matrices and fit a low-rank evolution operator using exact, optimized, or Hankel DMD.
- **Modal Decomposition:** Extract DMD eigenvalues, spatial modes, and projection weights, which encode frequencies, decay/growth rates, coherent structures, and modal amplitudes.
- **Physical Reconstruction:** Use the extracted modes to reconstruct dispersion, spectral density, IPR/localization diagnostics, and, when smooth momentum-space mode families are available, geometric and topological quantities.

This equation-based framework provides a clear microscopic interpretation, but its applicability depends on having an explicit and tractable model. In addition, direct access to the full phase structure of wavefunctions is often limited in experiment. These limitations motivate the use of Koopman-DMD as a complementary data-driven framework, in which band-related dynamical structure, localization signatures, and selected geometric or topological diagnostics are inferred from spatiotemporal measurements rather than derived from a known Hamiltonian.

## Section S2: Tight-Binding Models, Band Analysis, and Topological Phase Transitions from Hamiltonian FBD

In this section, we provide the explicit tight-binding Hamiltonians in real space and derive their corresponding momentum-space Bloch Hamiltonians. This theoretical framework underpins the band structure analysis, topological invariant calculations, and localization phenomena discussed in the main text.

### A. Anderson Localization and AAH Model

The 1D tight-binding model with nearest-neighbor hopping $t$ serves as the baseline for studying localization. The clean real-space Hamiltonian is $H_{\text{clean}} = -t \sum_i \left(c_i^\dagger c_{i+1} + \text{h.c.}\right)$.

To investigate the Anderson localization, a random on-site potential is introduced:

$$H_{\text{Anderson}} = H_{\text{clean}} + \sum_i U_i c_i^\dagger c_i \tag{S10}$$

where $U_i$ is uniformly distributed within the range $[-W/2, W/2]$. According to scaling theory, in a purely 1D system, any infinitesimally small disorder $W > 0$ leads to the exponential localization of all single-particle eigenstates, rendering the system an Anderson insulator.

In contrast, the Aubry-André-Harper (AAH) model introduces a deterministic, incommensurate quasi-periodic potential:

$$H_{\text{AAH}} = H_{\text{clean}} + \sum_i V \cos(2\pi\alpha i + \phi)\, c_i^\dagger c_i \tag{S11}$$

where $\alpha$ is typically an irrational number (e.g., the golden ratio). Unlike the purely random Anderson model, the 1D AAH model exhibits a sharp localization-to-delocalization phase transition. All eigenstates are extended when the potential strength $V < 2t$, and all states become exponentially localized when $V > 2t$, providing a critical paradigm bridging clean crystals and fully disordered systems.

By feeding the spatiotemporal data into our DMD framework, we reconstruct the energy spectrum, the density of states (DOS), and its projected Hofstadter-butterfly structure as functions of the quasi-periodic potential strength $V$and phase $\phi$. The 1D DOS reconstructed from DMD matches the exact result well across the transition (Fig. S1a), while the 2D DOS as a function of energy and quasi-periodic phase further resolves the underlying spectrum. The spatial profiles of the DMD modes clearly display extended behavior in the weak regime ($V = 0.5$, Fig. S1c) and localized behavior in the strong regime ($V = 4.0$, Fig.S1d). This tendency is also reflected in Fig. S1e, in which the localization length decreases with respect to $V$ (Fig. S1e). At fixed $V = 2.0$, the DMD-extracted spectrum reproduces the characteristic band structure as a function of $\phi$ with the IPR colored, indicating that these in-gap modes are more highly localized than the bulk (Fig. S1f). As compared in Fig. S1g and S1h, the projected Hofstadter-butterfly spectrum extracted from both the Hamiltonian and DMD shows consistent overall structures. Crucially, we can pinpoint the phase transition by averaging the ground-state IPR and the quantum metric over the phase $\phi$ from 0 to $2\pi$. As the potential strength approaches the critical point $V \approx 2$, both the average IPR and the quantum metric exhibit a concurrent jump, perfectly demarcating the transition boundary.

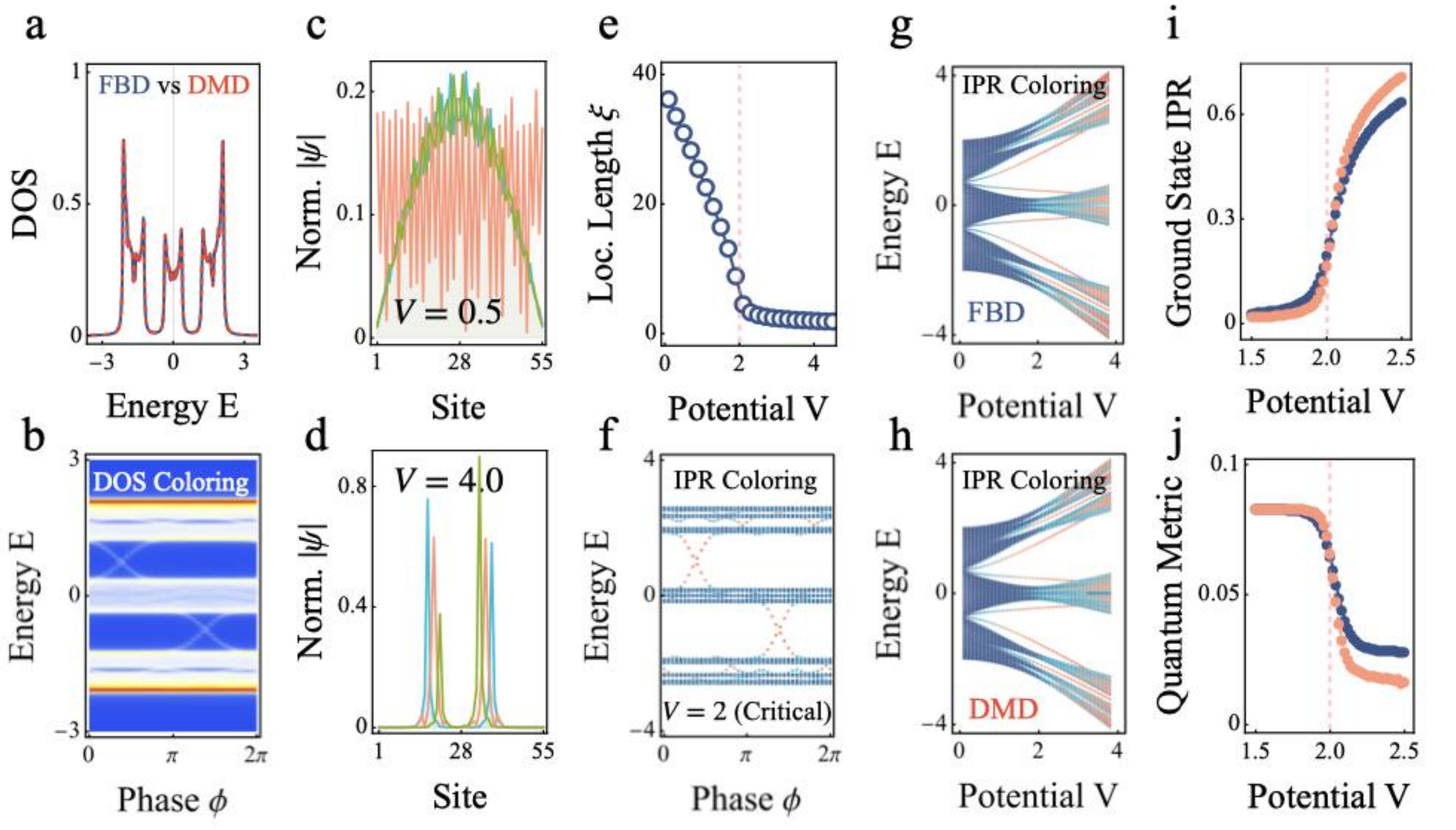

**Figure S1 | DMD analysis of the Aubry-André-Harper (AAH) model and its phase transition.** **(a)** Comparison of 1D density of states (DOS) between the exact Hamiltonian calculation and the DMD reconstruction, showing excellent agreement across the transition. **(b)** 2D DOS as a function of the energy and quasiperiodic phase ϕ, obtained from DMD. **(c, d)** Representative spatial profiles of DMD modes in the extended ($V = 0.5$) and localized ($V = 4.0$) regimes. **(e)** The localization length of DMD modes changing with potential $V$. **(f)** DMD-extracted energy spectrum at critical $V = 2$, plotted as a function of the quasi-periodic phase ϕ.The colormap corresponds to IPR. **(g, h)** The projected Hofstadter-butterfly spectrum extracted via Hamiltonian and DMD, with the coloring indicating a transition from the extended to localized. **(i, j)** Average IPR and quantum metric as a function of $V$. Both quantities exhibit a sharp, consistent increase near $V = 2$, while the inset shows the corresponding drop in localization length.

**B. Su-Schrieffer-Heeger (SSH) Model**

The 1D SSH model describes spinless fermions on a dimerized lattice with two sublattices (A and B) per unit cell. The real-space tight-binding Hamiltonian, featuring alternating intra-cell hopping $v$ and inter-cell hopping $w$, is given by:

$$H_{\mathrm{SSH}} = \sum_i \left( v c_{A,i}^{\dagger} c_{B,i} + w c_{A,i+1}^{\dagger} c_{B,i} + \mathrm{h.c.} \right) \tag{S12}$$

Applying the Fourier transform $c_{\alpha,i} = \frac{1}{\sqrt{N}} \sum_k e^{ikR_i} c_{\alpha,k}$ yields the momentum-space Bloch Hamiltonian:

$$H_{\mathrm{SSH}}(k) = \begin{pmatrix} 0 & v + we^{-ik} \\ v + we^{ik} & 0 \end{pmatrix} = d_x(k)\sigma_x + d_y(k)\sigma_y \quad \text{(S13)}$$

where $\sigma_{x,y}$ are the Pauli matrices, and the vector components are $d_x(k) = v + w\cos k$ and $d_y(k) = w \sin k$. The energy dispersion exhibits two symmetric bands:

$$E(k) = \pm\sqrt{v^2 + w^2 + 2vw\cos k} \quad \text{(S14)}$$

The topological phase transition occurs at $v = w$, where the bandgap closes at $k = \pi$. For $|v| < |w|$, the winding number is $\nu = 1$, corresponding to a non-trivial topological phase with zero-energy edge states under open boundary conditions.

### C. Non-Hermitian SSH Model with Asymmetric Pumping

To study the non-Hermitian skin effect (NHSE), we introduce a Hatano-Nelson-type non-Hermitian hopping $\gamma$ to the intra-cell couplings. The right-hopping amplitude becomes $v + \gamma$, and the left-hopping amplitude becomes $v - \gamma$:

$$H_{\text{NH-SSH}} = \sum_i \left[ (v+\gamma) c_{B,i}^\dagger c_{A,i} + (v-\gamma) c_{A,i}^\dagger c_{B,i} + w c_{A,i+1}^\dagger c_{B,i} + w c_{B,i}^\dagger c_{A,i+1} \right] \quad \text{(S15)}$$

Transforming to $k$-space, the Bloch Hamiltonian takes the non-Hermitian form:

$$H_{\text{NH-SSH}}(k) = \begin{pmatrix} 0 & (v-\gamma) + we^{-ik} \\ (v+\gamma) + we^{ik} & 0 \end{pmatrix} \quad \text{(S16)}$$

Diagonalizing this matrix yields a complex energy spectrum:

$$E(k) = \pm\sqrt{(v + w\cos k)^2 + (w \sin k - i\gamma)^2} \quad \text{(S17)}$$

The imaginary component leads to the formation of a point gap in the complex energy plane under periodic boundary conditions, which topologically dictates the macroscopic accumulation of eigenstates at the boundaries (NHSE). Besides, to investigate the Anderson localization, a random on-site potential is introduced: $H_{\text{NH-SSH}} + \sum_i \left( U_{A,i} c_{A,i}^\dagger c_{A,i} + U_{B,i} c_{B,i}^\dagger c_{B,i} \right)$, which allows us to explore the competition mechanism among Anderson localization, NHSE and topological localization.

### D. Floquet SSH Model: Periodically Driven Dynamics and Anomalous Topological Phases

To extend the analysis to the non-equilibrium regime, we consider a continuously driven Floquet SSH model. We express the time-dependent intra-cell coupling $v(t)$ and inter-cell coupling $w(t)$ by explicitly separating the mean hopping strength ($\kappa_0$), the static dimerization ($\delta\kappa_0$), and the periodically driven dimerization ($\delta\kappa$): $v(t) = \kappa_0 + \delta\kappa_0 + \delta\kappa\cos(\omega t)$, $w(t) = \kappa_0 - \delta\kappa_0 -$ $\delta\kappa\cos(\omega t)$. In momentum space, the time-dependent Bloch Hamiltonian reads $H(k,t) = d_x(k,t)\sigma_x + d_y(k,t)\sigma_y$, where:

$$
\begin{aligned}
d_x(k,t) &= [\kappa_0 + \delta\kappa_0 + \delta\kappa\cos(\omega t)] + [\kappa_0 - \delta\kappa_0 - \delta\kappa\cos(\omega t)]\cos k \\
d_y(k,t) &= [\kappa_0 - \delta\kappa_0 - \delta\kappa\cos(\omega t)]\sin k
\end{aligned}
\tag{S18}
$$

The dynamics are governed by the Floquet time-evolution operator $U(k,T) = \mathcal{T}\exp\left(-i\int_0^T H(k,t)dt\right)$, where $T = 2\pi/\omega$. The quasi-energies $\varepsilon(k)$ are extracted from the effective Floquet Hamiltonian $H_{\text{eff}}(k) = \frac{i}{T}\ln U\,(k,T)$ and are restricted to the Floquet Brillouin zone $[-\omega/2, \omega/2]$.

This driven system exhibits profound competition between the static topological mass ($\delta\kappa_0$) and the dynamically generated Floquet mass ($\delta\kappa$). A sufficiently strong driving field can optically force a topologically trivial static insulator into a Floquet topological phase by closing and reopening the bandgap at the zone center ($\varepsilon = 0$). Simultaneously, single-photon resonances bridge the bands at the zone boundaries, dynamically tearing open new bandgaps at $\varepsilon = \pm\omega/2$. The overall band topology is characterized by the dynamic invariant $\nu_{\text{band}} = \nu_0 - \nu_\pi$, guaranteeing the emergence of anomalous topological 0-modes and $\pi$-modes.

**Band Topology and Quantum Metric Extraction of Floquet systems.** The topological properties of 1D crystalline systems are fundamentally governed by their quantum geometry, specifically the quantum metric and the geometric phase (Zak phase). While extracting these geometric quantities typically requires access to global momentum-space eigenstates, we demonstrate that they can be directly reconstructed from real-space wavepacket dynamics using DMD.

For the static SSH model, we apply DMD to spatiotemporal data excited by random initial distributions (Fig. S2a1, a2). The extracted modes accurately reproduce the band structure in both the topologically trivial and nontrivial phases (b1, b2). The corresponding link variables vary smoothly across the Brillouin zone (c1, c2), indicating that the extracted modes are continuously connected in momentum space and free from artificial mode mixing. From these DMD modes, we compute the quantum metric $g_{kk}$ and reconstruct the Berry connection. In the trivial phase, $g_{kk}$ remains small over the Brillouin zone, whereas in the nontrivial phase it is strongly enhanced near the zone boundaries (d1, d2). Furthermore, by integrating the extracted Berry connection (e1, e2), we obtain the cumulative Zak phase, which sharply distinguishes the trivial (0) and nontrivial (π) phases (f1, f2).

This DMD-based geometric extraction naturally extends to periodically driven systems. For the Floquet SSH model, we consider two driven phases: one supporting only a π-mode and another supporting both 0- and π-modes (Fig. S2a3, a4, b3, b4). The DMD-extracted Floquet mode overlaps confirm smooth band connectivity (c3, c4), enabling the evaluation of the Floquet quantum metric. The DMD-extracted quantum metric is enhanced in the $\pi$-only phase but becomes much smaller when the 0 and $\pi$ modes coexist (d3, d4). The resulting Berry connection and cumulative Zak phase further distinguish these two driven phases (e3, e4, f3, f4). In particular, the band Zak phase returns to zero in the coexistence phase even though both 0 and $\pi$ modes are present, reflecting the anomalous Floquet relation $\nu_{band} = \nu_0 - \nu_\pi$ rather than a single static-like band invariant. These results show that DMD is not only a tool for band reconstruction, but a comprehensive quantum-geometric probe capable of faithfully reconstructing the underlying Fubini–Study metric and topological invariants purely from observables.

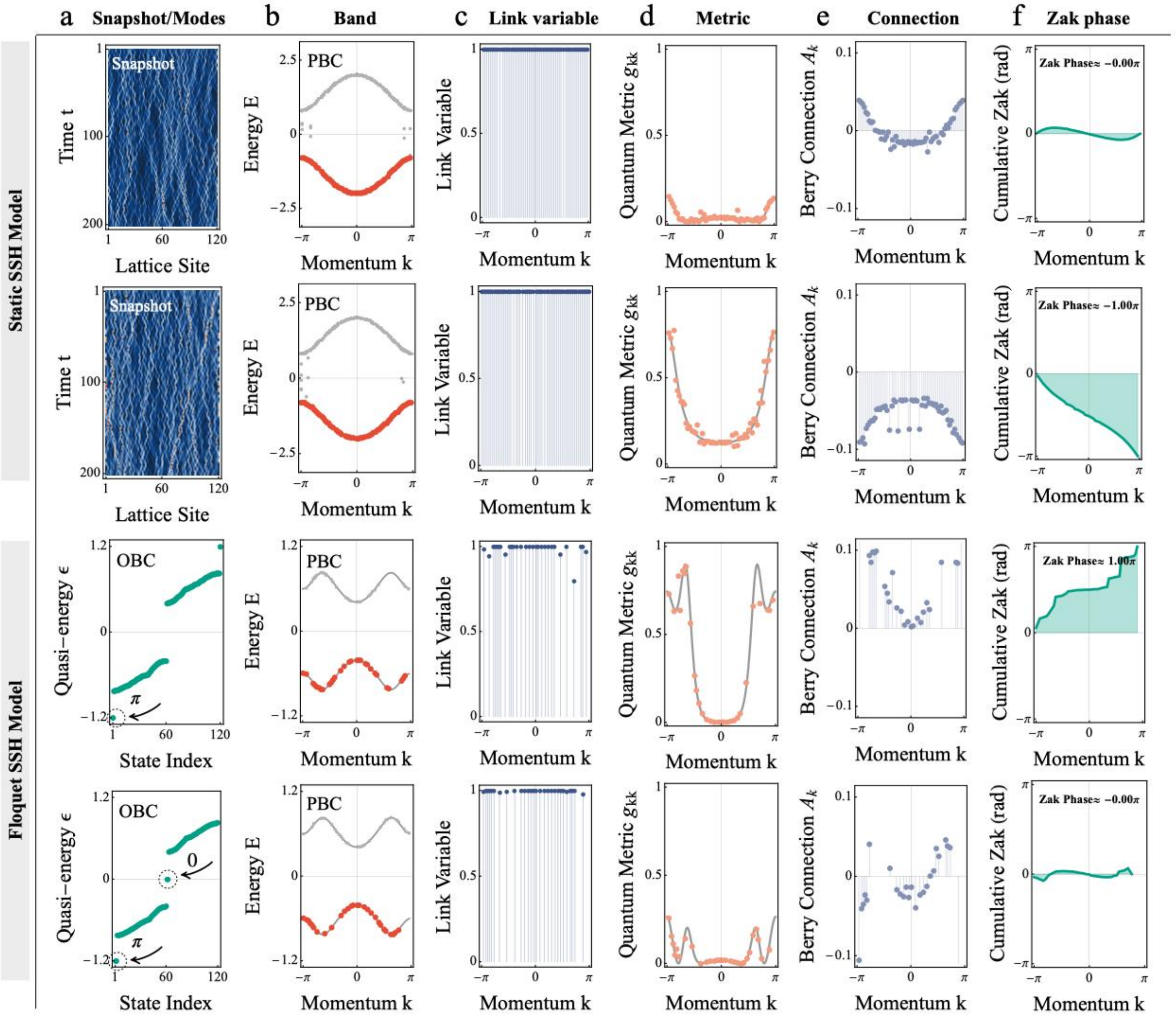

**Figure S2 | Data-driven band topology and quantum metric of the static and Floquet SSH models.** The first two rows show the trivial and nontrivial phases of static SSH model, while the last two rows show two driven SSH phases, one supporting only a $\pi$ mode and the other supporting coexisting 0- and $\pi$-modes. In each row, the six columns from left to right give: **(a)** The spatiotemporal snapshots for the static case in first two rows, and quasienergy distributions for the Floquet case in last two rows. **(b)** DMD-extracted band structures or quasienergy bands, with the band highlighted in red for extracting band topology. **(c)** Link variable varies smoothing in momentum space, indicating no artificial mode mixing. **(d)** Extracted quantum metric $g_{kk}$ (salmon-colored dots) compared with theoretical values (gray curves). In trivial phase of static case, $g_{kk}$ is much lower than the topological phase. For Floquet SSH, $g_{kk}$ is high for only pi mode existence. While it becomes small when both zero and pi mode present. **(e, f)** Reconstructed Berry connection and cumulative Zak phase. In particular, the Zak phase become zero when both zero and pi mode coexist, indicating the anomalous Floquet phases for driven system: $\nu_{band} = \nu_0 - \nu_\pi$.

### E. 2D SSH Model (Higher-Order Topological Insulator and Corner states)

The 2D SSH model on a square lattice is constructed by generalizing the 1D dimerized couplings to both spatial dimensions ( $x$ and $y$ ). The system consists of a square Bravais lattice where each unit cell (indexed by the lattice vector $\boldsymbol{R} = m\hat{x} + n\hat{y}$) contains four sublattices (A, B, C, D).

We define the intra-cell hopping amplitudes as $v_x$ and $v_y$, and the inter-cell hopping amplitudes as $w_x$ and $w_y$. The real-space tight-binding Hamiltonian is explicitly written as the sum of hoppings along the $x$ and $y$ directions $H_{2\mathrm{D}} = H_x + H_y$, where:

$$\begin{aligned} H_x &= \sum_{\boldsymbol{R}}\left[v_x\left(c_{A,\boldsymbol{R}}^\dagger c_{B,\boldsymbol{R}} + c_{C,\boldsymbol{R}}^\dagger c_{D,\boldsymbol{R}}\right) + w_x\left(c_{B,\boldsymbol{R}}^\dagger c_{A,\boldsymbol{R}+\hat{x}} + c_{D,\boldsymbol{R}}^\dagger c_{C,\boldsymbol{R}+\hat{x}}\right)\right] + \text{h.c.} \\ H_y &= \sum_{\boldsymbol{R}}\left[v_y\left(c_{A,\boldsymbol{R}}^\dagger c_{C,\boldsymbol{R}} + c_{B,\boldsymbol{R}}^\dagger c_{D,\boldsymbol{R}}\right) + w_y\left(c_{C,\boldsymbol{R}}^\dagger c_{A,\boldsymbol{R}+\hat{y}} + c_{D,\boldsymbol{R}}^\dagger c_{B,\boldsymbol{R}+\hat{y}}\right)\right] + \text{h.c.} \end{aligned} \tag{S19}$$

Applying the Fourier transform $c_{\alpha,\boldsymbol{R}} = \frac{1}{\sqrt{N}}\sum_{\boldsymbol{k}} e^{i\boldsymbol{k}\cdot\boldsymbol{R}} c_{\alpha,\boldsymbol{k}}$, the momentum-space Hamiltonian is expressed using the Kronecker product of two 1D SSH models:

$$H_{2\mathrm{D}}(\boldsymbol{k}) = H_{\mathrm{SSH},x}(k_x) \otimes I_2 + I_2 \otimes H_{\mathrm{SSH},y}(k_y) \tag{S20}$$

Explicitly, it takes the 4 $\times$ 4 form:

$$H_{2\mathrm{D}}(k_x, k_y) = \begin{pmatrix} 0 & h_x & h_y & 0 \\ h_x^* & 0 & 0 & h_y \\ h_y^* & 0 & 0 & -h_x \\ 0 & h_y^* & -h_x^* & 0 \end{pmatrix} \tag{S21}$$

where $h_{x,y} = v_{x,y} + w_{x,y}e^{-ik_{x,y}}$. While the bulk Chern number remains trivially zero, the non-trivial nested Wilson loop ensures the emergence of highly localized, zero-energy corner states under full open boundary conditions, characterizing it as a higher-order topological insulator (HOTI).

### F. The Honeycomb Lattice: From Graphene to the Haldane model

The honeycomb lattice is built from a triangular Bravais lattice with a two-site basis corresponding to sublattices A and B. Taking the lattice constant to be $a_0$, the three nearest-neighbor (NN) vectors connecting an A-site to its neighboring B-sites are:

$$\boldsymbol{\delta_1} = (0, -a_0), \boldsymbol{\delta_2} = \left(\frac{\sqrt{3}a_0}{2}, \frac{a_0}{2}\right), \boldsymbol{\delta_3} = \left(-\frac{\sqrt{3}a_0}{2}, \frac{a_0}{2}\right) \tag{S22}$$

**Graphene:** With only NN hopping amplitude $t$, the real-space tight-binding Hamiltonian is:

$$H_0 = -t\sum_{\boldsymbol{R}}\sum_{l=1}^{3}\left(c_{A,\boldsymbol{R}}^\dagger c_{B,\boldsymbol{R}+\boldsymbol{\delta}_l} + \text{h.c.}\right) \tag{S23}$$

In momentum space, using $\Psi_{\boldsymbol{k}} = \left(c_{A,\boldsymbol{k}}, c_{B,\boldsymbol{k}}\right)^T$, the phase accumulation from the three NN vectors yields:

$$H_0(\boldsymbol{k}) = \begin{pmatrix} 0 & -tf(\boldsymbol{k}) \\ -tf^*(\boldsymbol{k}) & 0 \end{pmatrix}, f(\boldsymbol{k}) = \sum_{l=1}^{3} e^{i\boldsymbol{k}\cdot\boldsymbol{\delta}_l} \tag{S24}$$

Due to the preservation of both Time-Reversal Symmetry (TRS) and Inversion Symmetry (IS), $f(\boldsymbol{k}) = 0$ at the Brillouin zone corners, forming gapless Dirac cones at the $K$ and $K'$ valleys.

**The Haldane Model:** To realize the quantum anomalous Hall (QAH) effect, the Haldane model introduces two symmetry-breaking perturbations to the real-space lattice:

$$H_{\text{Haldane}} = H_0 + M\sum_{\boldsymbol{R}}\left(c^\dagger_{A,\boldsymbol{R}}c_{A,\boldsymbol{R}} - c^\dagger_{B,\boldsymbol{R}}c_{B,\boldsymbol{R}}\right) - t_2\sum_{\langle\langle i,j\rangle\rangle} e^{i\phi_{ij}} c_i^\dagger c_j \tag{S25}$$

The first term is a staggered sublattice potential $M$ which breaks inversion symmetry. The second term is a complex next-nearest-neighbor (NNN) hopping connecting atoms of the same sublattice, breaking TRS. In $k$ -space, these terms generate an effective mass $d_z(\boldsymbol{k}) = M - 2t_2\sin\phi\sum_{j=1}^{3}\sin\left(\boldsymbol{k}\cdot\boldsymbol{v_j}\right)$ (where $\boldsymbol{v_j}$ are the NNN lattice vectors). The system undergoes a topological phase transition at $|M| = 3\sqrt{3}t_2|\sin\phi|$. When $|M| < 3\sqrt{3}t_2|\sin\phi|$, band inversion occurs at one valley, yielding a non-zero Chern number ( $C = \pm 1$ ) and chiral edge states.

**G. Frustrated Lattices and Flat Bands (Lieb & Kagome)**

In geometrically frustrated lattices, destructive quantum interference can quench the kinetic energy, giving rise to strictly flat bands where the energy $E(k)$ remains constant across the Brillouin zone. The resulting macroscopic density of states makes these systems highly sensitive to interactions, serving as fertile ground for strong correlation phenomena such as heavy fermion superconductivity. Because the kinetic energy is quenched, the physical properties and topological responses of these flat bands are entirely dictated by their underlying quantum geometry. Geometric frustration in specific lattice geometries leads to destructive quantum interference, completely quenching the kinetic energy and resulting in perfectly flat bands. We adopt a periodic gauge where the intra-cell spatial coordinates are absorbed into the Bloch basis.

**Lieb Lattice:** The Lieb lattice is an edge-centered square lattice. The unit cell at $\boldsymbol{R} = (m,n)$ contains three atoms: A at $(m,n)$, B at $(m+1/2,n)$, and C at $(m,n+1/2)$. With nearest-neighbor hopping $t$, the real-space Hamiltonian is:

$$H_{\text{Lieb}} = -t\sum_{m,n}\left[\left(c^\dagger_{A,m,n}c_{B,m,n} + c^\dagger_{A,m+1,n}c_{B,m,n}\right) + \left(c^\dagger_{A,m,n}c_{C,m,n} + c^\dagger_{A,m,n+1}c_{C,m,n}\right)\right] + \text{h.c.} \tag{S26}$$

Absorbing the spatial offset into the Fourier transform gives momentum-space Hamiltonian:

$$H_{\text{Lieb}}(\boldsymbol{k}) = \begin{pmatrix} 0 & -2t\cos(k_x/2) & -2t\cos\left(k_y/2\right) \\ -2t\cos(k_x/2) & 0 & 0 \\ -2t\cos\left(k_y/2\right) & 0 & 0 \end{pmatrix} \tag{S27}$$

Diagonalization reveals two dispersive bands and one strictly flat band $E_0(\boldsymbol{k}) = 0$, which intersect at the high-symmetry $M$ point.

**Kagome Lattice:** Composed of corner-sharing triangles, the Kagome unit cell contains three sublattices (A, B, C) located at the midpoints of the bonds of a simple hexagonal lattice. The real-space Hamiltonian sums over the nearest-neighbor bonds forming the triangles:

$$H_{\text{Kagome}} = -t \sum_{\langle i,j \rangle} c_i^\dagger c_j \tag{S28}$$

Defining the fractional momenta along the primitive reciprocal directions as $k_1 = \boldsymbol{k} \cdot \boldsymbol{a_1}/, k_2 = \boldsymbol{k} \cdot \boldsymbol{a_2}/$, and $k_3 = k_2 - k_1$, the mapping to momentum space yields:

$$H_{\text{Kagome}}(\boldsymbol{k}) = -2t \begin{pmatrix} 0 & \cos k_1 & \cos k_2 \\ \cos k_1 & 0 & \cos k_3 \\ \cos k_2 & \cos k_3 & 0 \end{pmatrix} \tag{S29}$$

The spectrum consists of two dispersive bands and a strictly flat band $E_0(\boldsymbol{k}) = 2t$. The completely quenched kinetic energy in these macroscopic degenerate flat bands implies that their physical behavior and topological responses are entirely dominated by the quantum geometric tensor.

To capture these exotic states, we apply our Koopman-DMD framework to two prototypical flat-band models: the Lieb and Kagome lattices. Starting from their real-space lattice geometries (Fig. S3a), the DMD algorithm reconstructs the dispersive bands together with the characteristic flat bands, clearly resolving the band structures at high-symmetry points (S3b). The reconstructed quantum metric components exhibit good agreement with the theoretical surfaces for both lattices (S3c–e). In addition, the reconstructed Berry curvature reveals distinct momentum-space structures associated with the flat-band-touching regions (S3f). These results confirm that our data-driven approach can faithfully reconstruct the distinct band-resolved quantum geometries of 2D flat-band lattices directly from spatiotemporal dynamics.

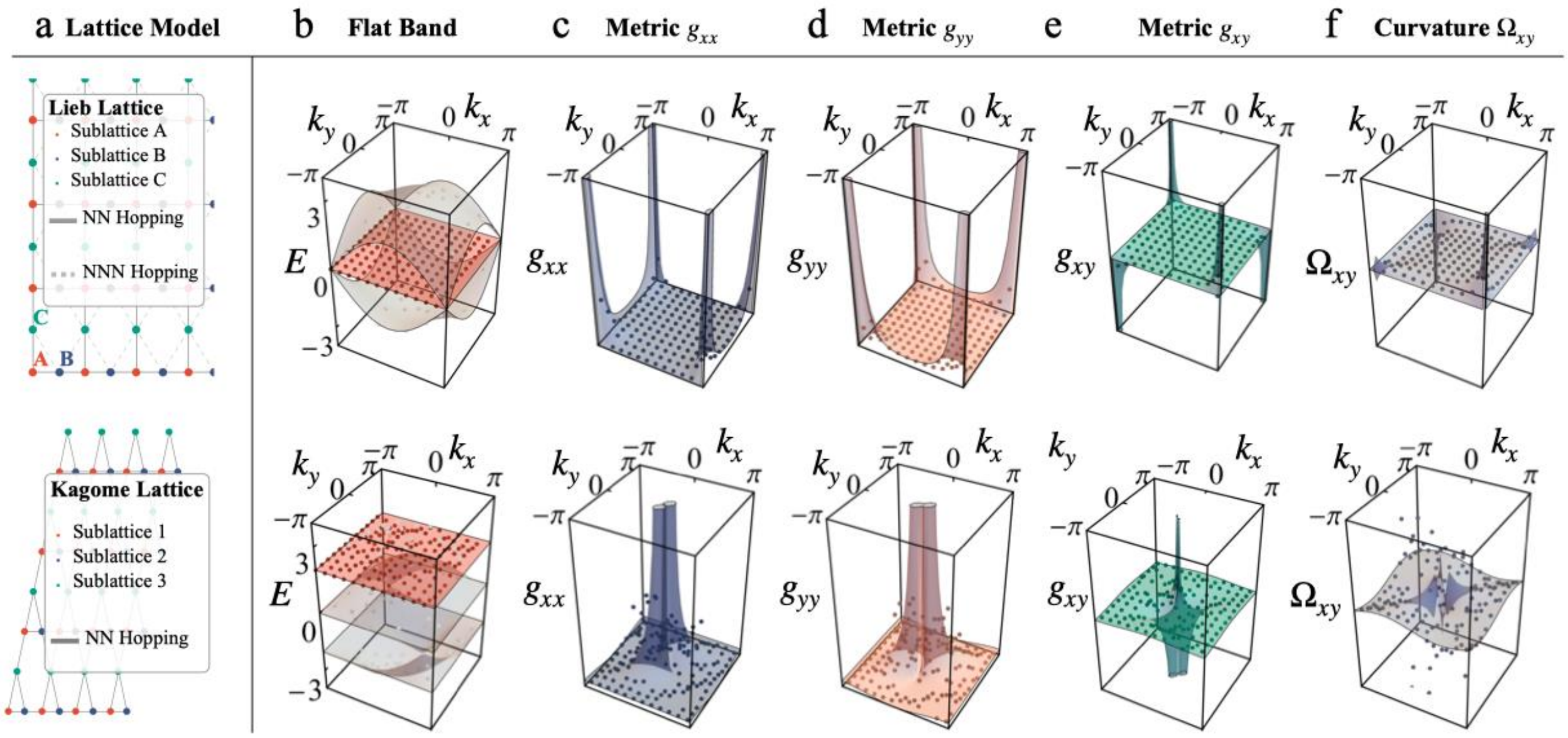

**Figure S3 | DMD analysis of the flat-band structure and quantum geometry in the Lieb and Kagome lattices.** Each column is labeled **(a)–(f)** from left to right. The upper panels correspond to the Lieb lattice, and the lower panels to the Kagome lattice. **(a)** Schematic illustrations of the Lieb and Kagome lattices. **(b)** Reconstructed Bloch-band structures, highlighting the DMD-extracted modes selected for the quantum-geometric analysis. **(c)–(e)** Comparison between the DMD-extracted quantum metric components (black markers) and the exact theoretical surfaces (colored surfaces). The distinct momentum-space geometric structures reflect the unique flat-band properties of the two frustrated lattices. **(f)** Reconstructed Berry curvature, revealing the characteristic momentum-space distribution associated with the flat-band-touching singularities.

## Section S3: Readout noise and reconstruction robustness

In practical quantum devices, coupling to the environment, fluctuations in control parameters, and noise in the measurement chain can all affect experimental results. Extracting the relevant physical information from noisy spatiotemporal data is therefore essential. The reduced representation provided by Koopman DMD allows rank truncation to suppress noise contributions. Here, we use the SSH model to assess reconstruction robustness: its band dispersion tests the accuracy of the extracted modes, while its robust topological properties provide a complementary test. We first examine how noise affects the input data and DMD reconstruction, then investigate the effectiveness and limitations of truncation in suppressing noise.

We model readout noise by adding Gaussian noise to the complex spatiotemporal snapshots. Let $X_0$ denote the noise-free snapshot matrix, with spatial dimension $N$ and $M$ snapshots. Readout noise is added after the evolution has been computed:

$$X_\varepsilon = X_0 + \frac{\varepsilon\sigma_0}{\sqrt{2}}(\varXi + iZ), \quad \sigma_0 = \frac{\|X_0\|_\mathrm{F}}{\sqrt{NM}} \tag{S30}$$

The entries of $\varXi$ and $Z$ are mutually independent standard normal random variables. Figure S4 illustrates the effect of noise on the spatiotemporal data: the noise-free structure in Fig. S4a1 becomes increasingly obscured as the noise strength increases in Figs. S4b1–d1. We quantify the input error relative to the noise-free data as

$$e_\mathrm{in} = \frac{\|X_\varepsilon - X_0\|_\mathrm{F}}{\|X_0\|_\mathrm{F}} \tag{S31}$$

where $\|\cdot\|_F$ denotes the Frobenius norm. The input error $e_{in}$ increases approximately linearly with the noise strength $\varepsilon$.

We then reconstruct the spatiotemporal data using DMD:

$$\hat{x}_m = \varPhi \varLambda^m b, \quad b = \varPhi^+ x_{\varepsilon,0} \tag{S32}$$

Here, $\varPhi$ is the mode matrix, $\varLambda$ is the diagonal matrix of eigenvalues, $\varPhi^+$ is the pseudoinverse of $\varPhi$, and $x_{\varepsilon,0}$ is the noisy initial snapshot. As the noise strength increases, the reconstructed dynamics exhibit progressively stronger decay because fitting noisy snapshots introduces a decay bias into the DMD eigenvalues. Figure S4e2 shows the reconstruction error relative to the noise-free data in Fig. S4a1, defined as

$$e_X = \frac{\|\hat{X} - X_0\|_\mathrm{F}}{\|X_0\|_\mathrm{F}} \tag{S33}$$

where $\hat{X}$ contains the reconstructed snapshots and $X_0$ contains the noise-free reference snapshots. The reconstruction error increases with noise strength and eventually reaches a maximum.

To examine how noise affects truncation, we perform SVD on the snapshot matrix used for DMD fitting and consider the relative weights of its singular values. For singular values $s_j$ arranged in descending order, the relative weight is defined as

$$w_j = \left(\frac{s_j}{s_1}\right)^2 \tag{S34}$$

Only singular directions satisfying $w_j > w_{cut}$ are retained, and their number defines the truncation rank $r$. In Fig. S4, we fix $w_{cut} = 10^{-4}$.

Figures S4a3–d3 show the relative-weight distributions at different noise strengths. As the noise increases, more noise contributions exceed this threshold, leading to a larger retained rank (Fig. S4e3). Therefore, the truncation threshold must be increased appropriately to reduce the number of retained noise contributions.

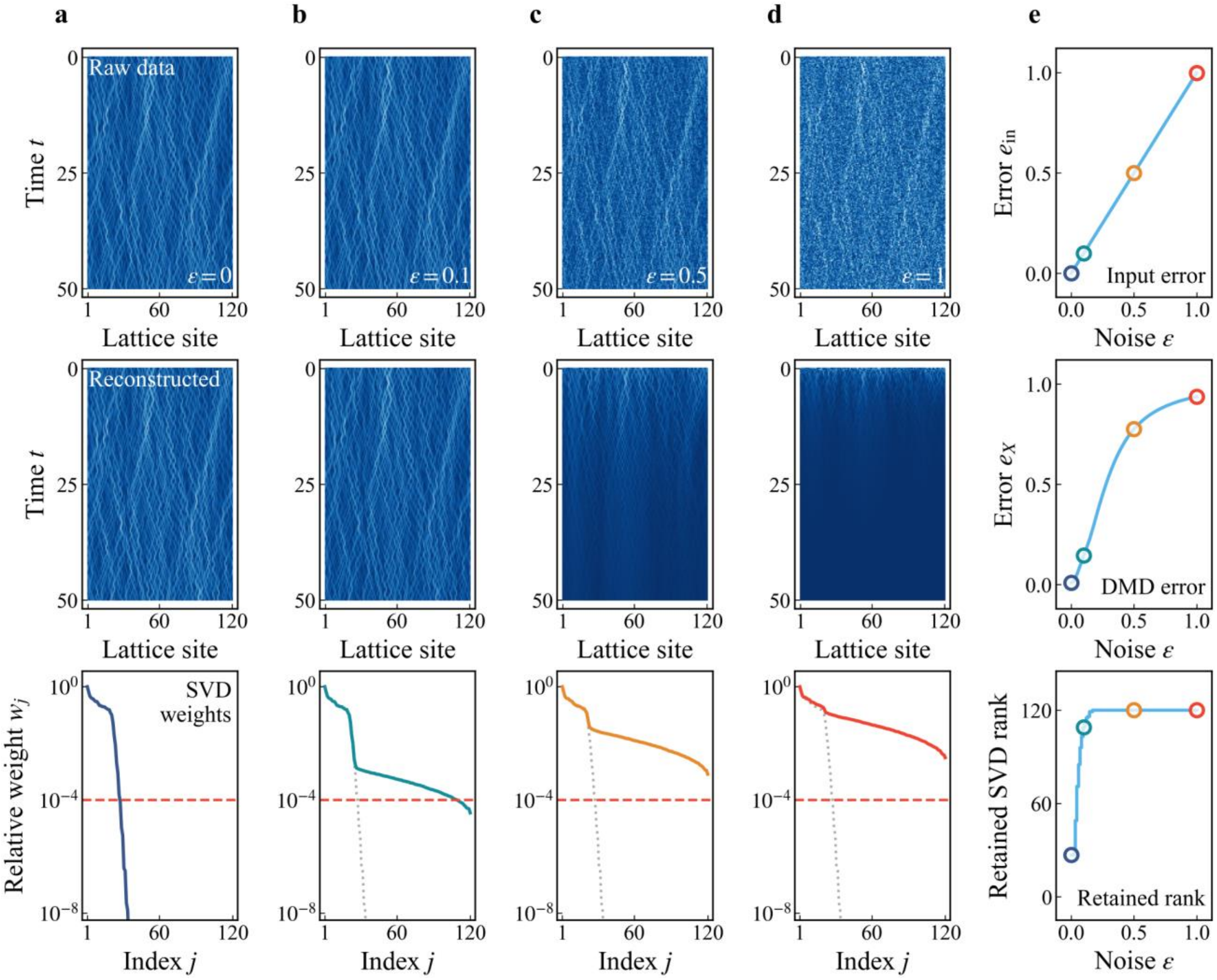

**Figure S4 | Effect of readout noise on DMD reconstruction at a fixed truncation threshold.** Panels a1–d1, a2–d2, and a3–d3 show the input spatiotemporal data, DMD reconstructions, and SVD relative-weight spectra, respectively. Columns a–d correspond to $\varepsilon = 0, 0.1, 0.5$, and 1. Panels e1–e3 show the input error, reconstruction error, and retained rank as functions of noise strength. Gray dotted lines indicate the noise-free weight reference, and red dashed lines mark the fixed truncation threshold.

We introduce a scoring scheme to evaluate the reliability of DMD reconstruction. Each component score ranges from 0 to 100, and the total score is defined as

$$S_{\text{total}} = 0.20S_X + 0.30S_E + 0.30S_L + 0.10S_g + 0.10S_\gamma \tag{S35}$$

We assign the largest weights, 0.30 each, to the band score $S_E$ and the link score $S_L$. Band reconstruction is the central objective of this work, while the link score characterizes the smooth variation of the reconstructed states with momentum. The spatiotemporal reconstruction score $S_X$ receives a weight of 0.20. The metric score $S_g$ and phase score $S_\gamma$, which evaluate quantities derived from the reconstructed states, each receive a weight of 0.10. Comparing the total score across truncation thresholds identifies the best threshold at a given noise strength (black curves in Figs. S5e1–e5). When band reconstruction is insufficient, the link, metric, and phase scores are not evaluated (Figs. S5a3–a5 and d3–d5); the corresponding thresholds are shaded red in the score plots.

The spatiotemporal reconstruction score is defined as

$$S_X = 100\max(0,1 - e_X) \tag{S36}$$

where $e_X$ is the reconstruction error defined above.

As fewer modes are retained, the reconstructed spatiotemporal data deviate further from the reference data (insets in Figs. S5a1–d1). Accordingly, the spatiotemporal reconstruction score decreases as the truncation threshold increases, as shown by the purple curve in Fig. S5e1.

For band reconstruction, we assess both agreement with the theoretical bands and coverage of those bands.

Each extracted point is classified using its minimum energy difference from the theoretical bands at the same momentum:

$$d_j = \min_{\alpha=\pm}\left|E_j - E_\alpha(k_j)\right|, \quad \delta_E = 0.05(|v| + |w|) \tag{S37}$$

Here, $(k_j, E_j)$ denotes the $j$th extracted point, and $E_\alpha(k)$ denotes the two theoretical bands. A point is classified as incorrect if $d_j > \delta_E$, and $p_{wrong}$ is the fraction of incorrect points. The band score is defined as

$$C_E = \frac{N_{\text{cov}}}{2N_k}, \quad S_E = 100\left(1 - p_{\text{wrong}}\right)P(C_E) \tag{S38}$$

Here, $N_k$ is the number of points in the full momentum grid, and $N_{cov}$ counts the distinct momentum–band-branch combinations covered by correctly identified points; repeated points in the same combination are counted only once. To prevent high scores based on only a few recovered points, we introduce a coverage factor for both the band and link scores:

$$P(C) = \min\left(1, \frac{C}{0.60}\right) \tag{S39}$$

No coverage penalty is applied at or above 60%; below this value, the score is reduced proportionally.

We evaluate the reconstructed geometry using the link, metric, and phase scores, which characterize overlaps between neighboring Bloch states, errors in the quantum metric, and deviations in the Zak phase, respectively.

The link score is based on the mean deviation of the overlap magnitude between neighboring normalized Bloch states from unity:

$$e_L = \frac{1}{N_L}\sum_\ell \left|1 - |U_\ell|\right|, \quad S_L = 100\max(0,1 - e_L)P(C_L) \tag{S40}$$

The closer the overlap magnitudes are to unity, the higher the link score. Here, $U_\ell$ is the overlap on link $\ell$, and $N_L$ is the number of links. The momentum coverage of the selected valence-band states is $C_L = N_{sel}/N_k$, where $N_{sel}$ is the number of distinct selected momenta.

The metric score measures the relative error between the reconstructed quantum metric and its theoretical value. Because the selected momenta may be unevenly spaced, we weight the error by the momentum interval $\Delta k_\ell$ of each link:

$$e_g = \sqrt{\frac{\sum_\ell \Delta k_\ell \left(g_\ell - g_\ell^{\mathrm{th}}\right)^2}{\sum_\ell \Delta k_\ell \left(g_\ell^{\mathrm{th}}\right)^2}}, \quad S_g = \frac{100}{1+e_g} \tag{S41}$$

Here, $g_\ell$ is the reconstructed quantum metric, and ${g_\ell}^{th}$ is its theoretical value at the midpoint of the corresponding link. A smaller error gives a score closer to 100.

Since phases are periodic modulo $2\pi$, the phase score uses the shortest phase difference between the reconstructed and theoretical Zak phases:

$$e_\gamma = \frac{\left|\mathrm{Arg}\left(e^{i(\gamma - \gamma_{\mathrm{th}})}\right)\right|}{\pi}, \quad S_\gamma = 100\left(1 - e_\gamma\right) \tag{S42}$$

Here, $\gamma$ and $\gamma_{th}$ are the reconstructed and theoretical Zak phases, respectively, and Arg denotes the principal argument.

To evaluate geometric properties only when band reconstruction is sufficiently reliable, the link, metric, and phase scores are calculated only if $S_E \geq 60$; otherwise, they are labeled error. The same label is used when fewer than two valence-band states are selected.

When the truncation threshold is too low, many noise modes are retained, causing extracted points to deviate from the theoretical bands (Fig. S5a2). Increasing the threshold suppresses these modes and improves the reconstruction of the bands and their geometric properties (Figs. S5b2 and c2). An excessively high threshold, however, also discards physical modes, leaving incomplete band coverage (Fig. S5d2). The band score therefore first increases and then decreases with the truncation threshold (Fig. S5e2).

Scanning the threshold thus identifies the best reconstruction at a fixed noise strength. We next examine the noise range over which truncation still allows the bands and related properties to be recovered.

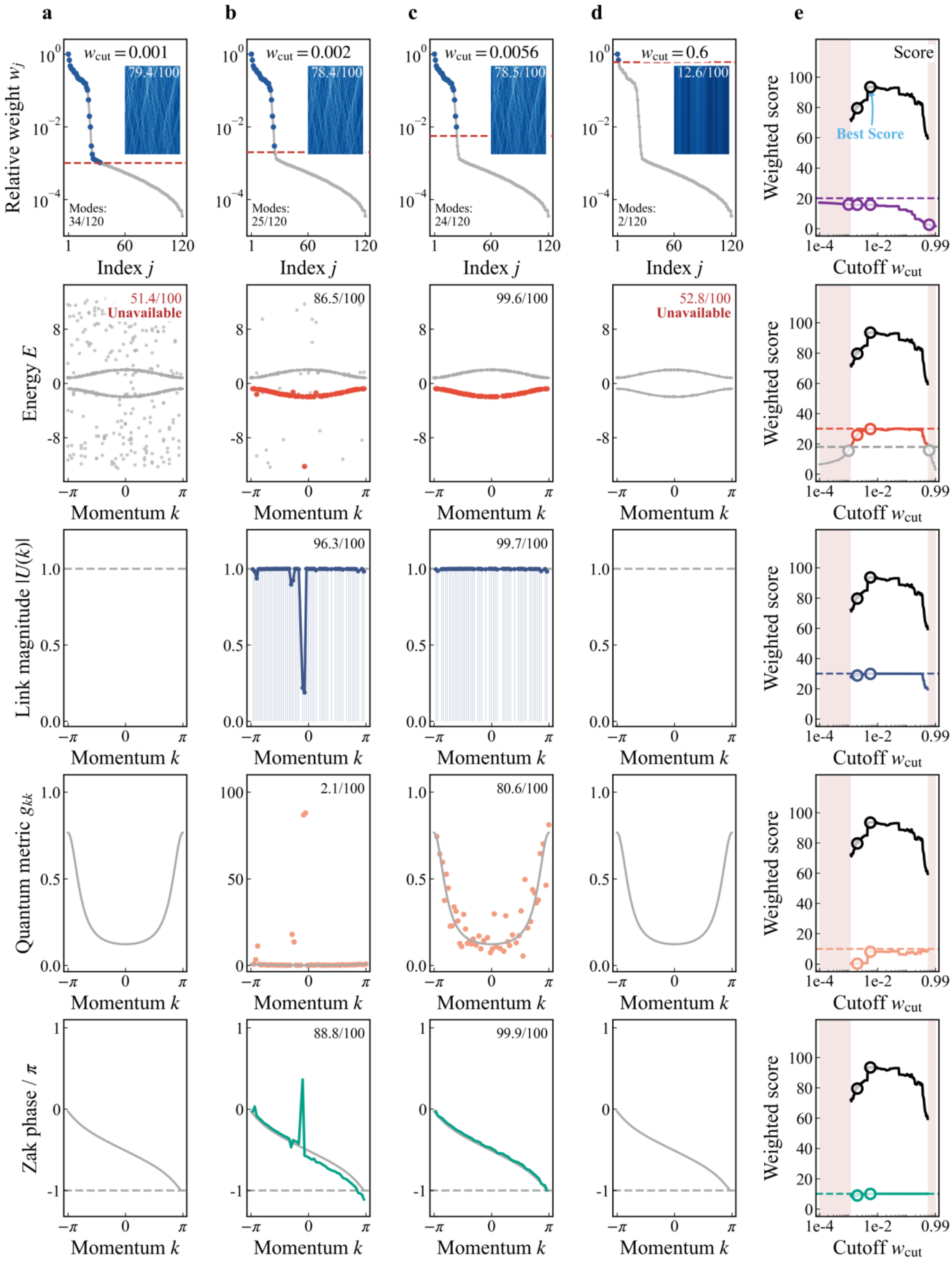
a
b
c
d
e
$w_{\rm cut} = 0.001$
$w_{\rm cut} = 0.002$
$w_{\rm cut} = 0.0056$
$w_{\rm cut} = 0.6$
79.4/100
78.4/100
78.5/100
12.6/100
Modes: 34/120
Modes: 25/120
Modes: 24/120
Modes: 2/120
Relative weight $w_j$
Index $j$
Score
Best Score
Weighted score
Cutoff $w_{\rm cut}$
51.4/100
Unavailable
86.5/100
99.6/100
52.8/100
Unavailable
Energy $E$
Momentum $k$
96.3/100
99.7/100
Link magnitude $|U(k)|$
2.1/100
80.6/100
Quantum metric $g_{kk}$
88.8/100
99.9/100
Zak phase / $\pi$

**Figure S5 | Truncation scan and reconstruction quality at a fixed readout noise strength of $\boldsymbol{\varepsilon = 0.1}$.** Columns a–d compare different relative-weight thresholds. Rows 1–5 show the SVD relative-weight spectra, bands, link magnitudes, quantum metric, and accumulated Zak phase, respectively. Column c uses the threshold that maximizes the total score in this scan. Insets in the SVD spectra show the spatiotemporal reconstructions; the number of retained directions is indicated. Panels e1–e5 show the weighted component scores and the total score (black), with the data score shown in purple. The light-blue circle and arrow in e1 mark the maximum total score. Gray curves indicate theoretical references. Red shading identifies thresholds that fail the eligibility criterion, and Unavailable indicates undefined geometric quantities.

Figure S6 compares reconstruction scores at the best threshold for each noise strength, with all component scores evaluated at the same selected threshold. As noise increases, the total score decreases, and the spatiotemporal reconstruction score drops rapidly, approaching zero at $\varepsilon \approx 0.9$. The band score remains close to its maximum at $\varepsilon \approx 0.6$ before gradually declining. In contrast, the link and Zak-phase scores remain high wherever they can be evaluated, while the quantum-metric score gradually decreases. These trends demonstrate that different physical quantities have different sensitivities to noise. The best threshold generally increases with noise strength to exclude noise contributions with growing relative weights. At $\varepsilon = 1.25$, no threshold satisfies the scoring criteria. To further illustrate the limitations of truncation under strong noise, Fig. S7 shows reconstructions at different thresholds for $\varepsilon = 1.2$.

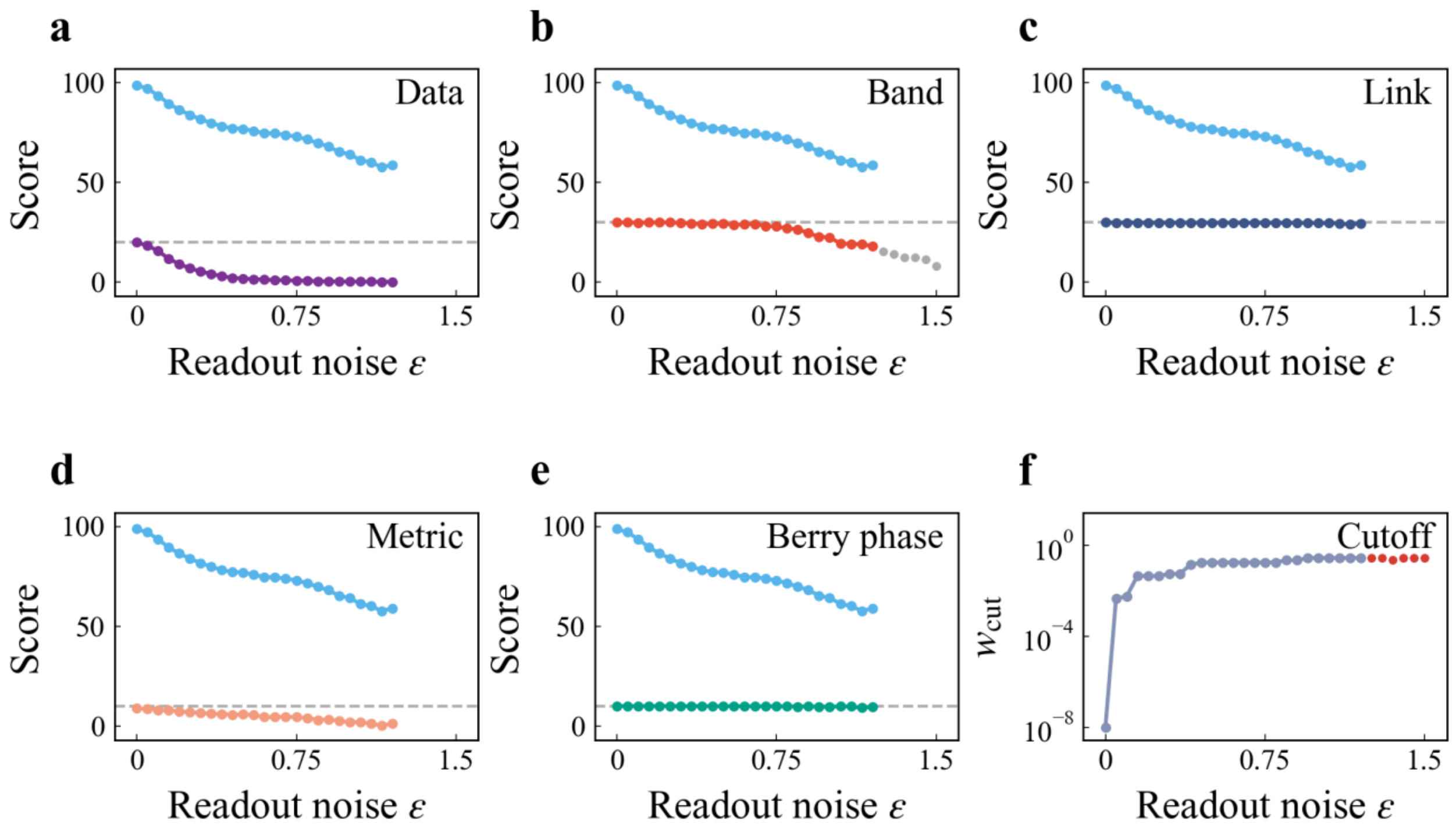

**Figure S6 | DMD reconstruction quality versus readout noise at the best truncation threshold.** Panels a–e show the weighted scores for spatiotemporal data, bands, links, quantum metric, and Berry phase, respectively, together with the total score. Panel f shows the selected relative-weight threshold at each noise strength. All results at a given noise strength use the same selected threshold. Red points denote fallback results when no eligible threshold is available; error indicates that the corresponding geometric quantities fail the eligibility criterion and are not evaluated.

At $\varepsilon = 1.2$, spatiotemporal reconstruction is poor because fitting the noisy snapshots introduces a strong decay bias, preventing the reconstructed dynamics from matching the reference data (insets in Figs. S7a1–d1). Noise-induced points persist in the reconstructed bands even at high thresholds (Fig. S7c2), and the retained points do not lie exactly on the theoretical bands even when few modes remain. Among the representative thresholds shown, only Fig. S7c2 has a band score above 60 and is therefore retained for geometric analysis. The link magnitudes remain close to unity (Fig. S7c3), but the quantum metric deviates substantially from theory (Fig. S7c4). The Zak phase nevertheless remains close to its theoretical value (Fig. S7c5). The highest total score in this scan is approximately 58.9.

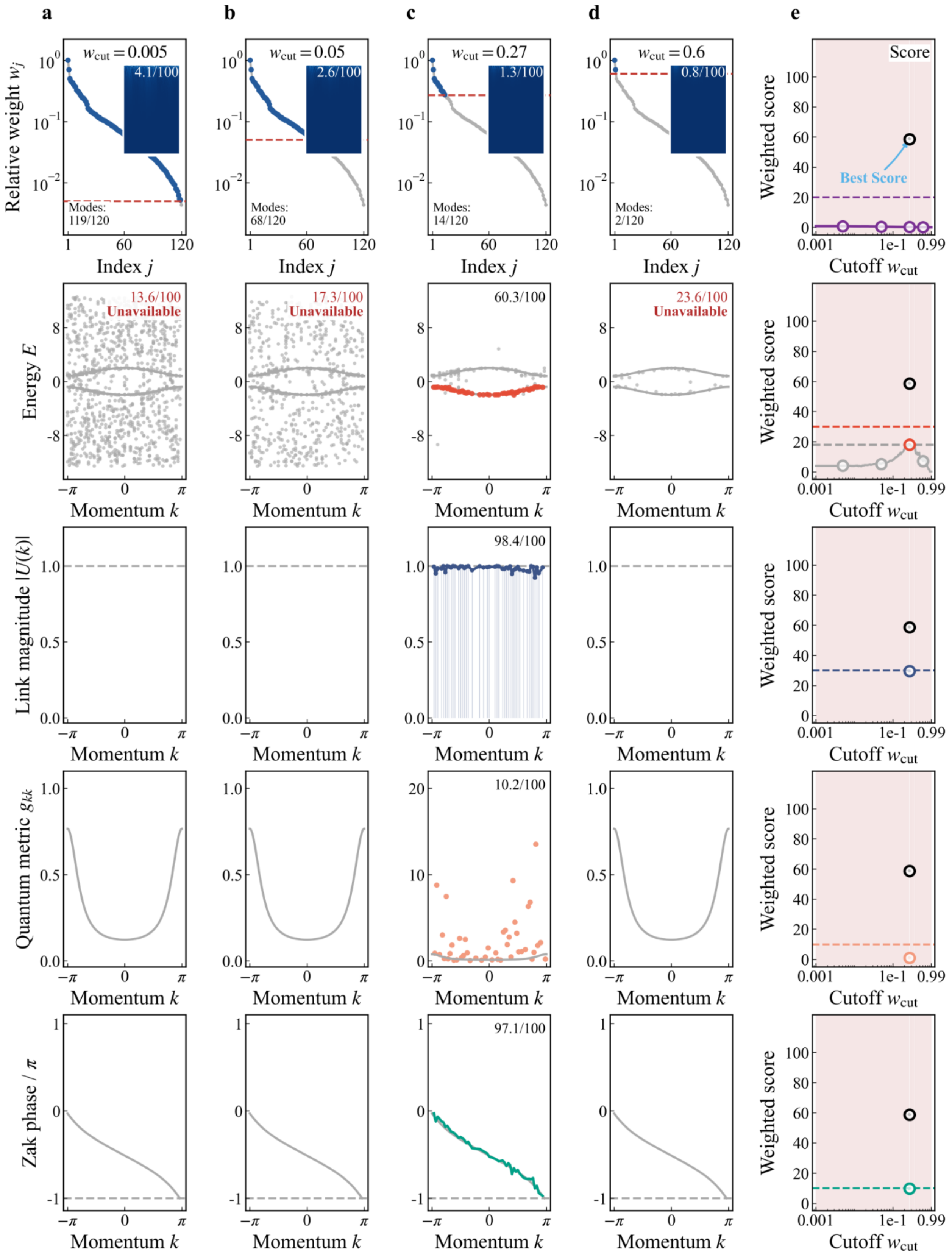
a
b
c
d
e
$w_{cut} = 0.005$
$w_{cut} = 0.05$
$w_{cut} = 0.27$
$w_{cut} = 0.6$
4.1/100
2.6/100
1.3/100
0.8/100
Modes: 119/120
Modes: 68/120
Modes: 14/120
Modes: 2/120
Relative weight $w_j$
Index $j$
Score
Best Score
Weighted score
Cutoff $w_{cut}$
13.6/100
Unavailable
17.3/100
Unavailable
60.3/100
23.6/100
Unavailable
Energy $E$
Momentum $k$
98.4/100
Link magnitude $|U(k)|$
10.2/100
Quantum metric $g_{kk}$
97.1/100
Zak phase / $\pi$

**Figure S7 | DMD reconstruction and truncation scan under strong readout noise at $\boldsymbol{\varepsilon = 1.2}$.** The layout and notation follow Fig. S5. Under strong noise, the reconstructed dynamics and quantum metric are substantially degraded, while the Zak phase remains close to its theoretical value at the selected threshold.

For the noisy SSH spatiotemporal data considered here, we select an appropriate truncation threshold to suppress noise while preserving as much physical information as possible. For noisy experimental data, however, theoretical reference values may be unavailable, preventing the best threshold from being identified through comparison with the DMD results as in this example. Selecting an appropriate truncation threshold may therefore require a case-by-case assessment of the reconstruction results.

**Section S4: Quasiparticle spectral function extracted from DMD.**

The modal frequencies and spatial structures extracted by DMD can be used to construct a data-driven spectral representation of the measured dynamics. If the modal frequencies $\omega_j$ are weighted by their projection amplitudes, the result is a set of discrete spectral peaks associated with the modes excited by the chosen initial condition. When the excitation has sufficiently broad spectral support, these peaks provide a useful approximation to the excited density of states or to a momentum-integrated spectral density. If the spatiotemporal field is further resolved in momentum space, or Fourier transformed from real space, one may also construct a momentum-resolved DMD spectral density $A(k,\omega)$.

From this perspective, DMD may be viewed as a form of finite-data system identification. The fitted linear evolution operator provides the reduced operator that best reproduces the observed dynamics within the chosen rank and time window. When the sampling interval is sufficiently small and the dynamics are approximately linear, one expects this operator to approximate the short-time propagator, so that its logarithm defines an effective generator. In the ideal limit of complete and noise-free data, this generator approaches the dominant spectral content of the true dynamics. In practice, finite time windows, noise, incomplete modal coverage, and damping all introduce reconstruction error. For this reason, the inferred operator should generally be interpreted as an effective reduced-order description rather than as a unique reconstruction of the microscopic Hamiltonian.

By contrast, within the Hamiltonian Floquet–Bloch decomposition (FBD), the spectral function is obtained from the retarded Green's function once the Hamiltonian is known. For a single-particle or mean-field system, $G^R(k,\omega) = [\omega + i0^+ - H(k)]^{-1}$, and the spectral function is $A(k,\omega) = -\frac{1}{\pi}\mathrm{Im}G^R(k,\omega)$. In noninteracting band models, $A(k,\omega)$ consists of δ-function peaks at the band energies $E_n(k)$, broadened in practice by finite lifetime or numerical smoothing. More generally, once the Hamiltonian is specified, the FBD framework provides a systematic route from microscopic dynamics to band dispersion, topology, and spectral response.

The single-particle spectral function $A(k,\omega)$ may be understood as the energy–momentum-resolved density of states for adding or removing a particle with momentum k. It is directly probed by ARPES (angle-resolved photoemission spectroscopy) on the occupied side and by inverse photoemission on the unoccupied side. In the standard Green's-function formulation, $A(k,\omega) = -\frac{1}{\pi}\mathrm{Im}G^R(k,\omega)$, with $G^R(k,\omega) = \frac{1}{\omega+\mu-\varepsilon_k-\Sigma^R(k,\omega)}$, where $\varepsilon_k$ is the bare or mean-field dispersion, and $\Sigma^R$ is the retarded self-energy. In the noninteracting limit, $\Sigma^R = 0$,, so the spectrum reduces to infinitely sharp bands. With interactions or scattering, $\Im(\Sigma^R) \neq 0$, and the δ-peaks broaden into finite-width quasiparticle resonances. Near an isolated pole, the spectral function takes the familiar form

$$A(k,\omega) \approx Z_k \frac{\Gamma_k/\pi}{(\omega-E_k)^2+\Gamma_k^2} + A_{\mathrm{incoh}} \quad \text{(S43)}$$

where $E_k$ is the renormalized dispersion, $Z_k$ the quasiparticle weight, and $\Gamma_{\boldsymbol{k}}$ the half-width, or inverse lifetime scale. Equivalently,

$$A(\boldsymbol{k},\omega) = \frac{1}{\pi}\frac{-\mathrm{Im}\Sigma^R(\boldsymbol{k},\omega)}{[\omega+\mu-\varepsilon_{\boldsymbol{k}}-\mathrm{Re}\Sigma^R(\boldsymbol{k},\omega)]^2+[\mathrm{Im}\Sigma^R(\boldsymbol{k},\omega)]^2} \tag{S44}$$

For a generic interacting lattice model, such as the Hubbard model, the exact spectral function is given by the Lehmann representation and therefore depends on many-body eigenstates: $A(k,\omega) = \sum_m \left|\langle m^{N+1}|c_k^\dagger|0^N\rangle\right|^2 \delta(\omega - (E_m^{N+1} - E_0^N)) + \sum_m |\langle m^{N-1}|c_{\boldsymbol{k}}\,|0^N\rangle|^2 \delta(\omega - (E_m^{N+1} - E_0^N))$ where the second term involves the N-1-particle sector. In practice, it is usually computed through one of several standard routes: direct diagonalization of the single-particle Hamiltonian in noninteracting or mean-field settings; exact diagonalization or Lanczos continued fractions for small interacting clusters; time-dependent DMRG or tensor-network methods in one dimension; quantum Monte Carlo combined with analytic continuation from imaginary time; or DMFT and cluster-DMFT methods, where the self-energy is first computed and then inserted into Dyson's equation. A general computational fallback is to evaluate the real-time retarded correlator $G_{ij}^R(t) = -i\theta(t)\left\langle\left[c_i(t), c_j^\dagger(0)\right]_+\right\rangle$, or its appropriate ordering convention, Fourier transform in space and time, and then take $-\frac{1}{\pi}\mathrm{Im}$.

Suppose now that the measured dataset is a **single-particle wavefunction** or an effective mean-field/BdG field evolving linearly, $\psi(t+\Delta t) \approx U\psi(t)$. In this case, Koopman–DMD provides a data-driven approximation to the eigenvalues and eigenmodes of U. One obtains $\psi(r,t) \approx \sum_l b_l\, \phi_l(r)\lambda_l^{t/\Delta t}$, where $\phi_l(r)$ are the DMD modes and $\lambda_l$ are the corresponding eigenvalues. Writing $\omega_l - i\gamma_l = \frac{i}{\Delta t}\log\lambda_l$, the real part gives the oscillation frequency and the imaginary part gives the decay rate. If the system is translationally invariant, one may analyze the data directly in momentum space by Fourier transforming $\psi_i(t) \to \psi(k,t)$ and performing DMD either at each momentum or on the full field followed by momentum projection of the extracted modes. The dominant $\omega(k)$ then reconstructs the dispersion $E(k) \simeq \hbar\omega(k)$, up to convention. For such linear single-particle dynamics, a natural DMD-based spectral function is

$$A_{\mathrm{DMD}}(\boldsymbol{k},\omega) \propto \sum_l |\langle \boldsymbol{k}|\phi_l\rangle|^2 \frac{\gamma_l}{(\omega-\omega_l)^2+\gamma_l^2}. \tag{S45}$$

Here, the linewidth $\gamma_l$ is determined by $|\lambda_l|$ and may reflect finite-time decay, dissipation, leakage, or effective damping. Conceptually, this construction is a structured alternative to a direct temporal Fourier transform: rather than simply resolving peaks in frequency space, DMD separates the signal into coherent damped exponentials with distinct spatial content. The resulting $A_{\mathrm{DMD}}(\boldsymbol{k},\omega)$ should therefore be interpreted as a **data-driven quasiparticle spectral function** for linearly evolving single-particle or mean-field dynamics.

The situation is different in a fully interacting many-body system. There, the exact spectral function is defined in terms of two-time correlation functions, especially the retarded Green's function, rather than from a single wavefunction snapshot. A single realization of a many-body state, or a single classical or mean-field field trajectory, is generally insufficient to reconstruct the full many-body $A(k,\omega)$. This distinction is essential when extending Koopman-DMD from the single-particle benchmarks of the main text to genuinely interacting systems.

However, if one can construct a time series of the single-particle Green's function, for example $G_{ij}(t) \equiv \langle c_i(t) c_j^\dagger(0) \rangle$ or $G_{ij}^R(t)$, then DMD becomes physically relevant again. In that case, the matrix-valued Green's-function time series, or an equivalent stacked representation, is used as the observable entering the Koopman-DMD analysis. Because DMD decomposes the data into damped oscillatory components, it can provide a compact reduced-order representation of the pole structure and effective damping encoded in G(t). After Fourier transformation, this yields a data-driven reconstruction of spectral features in $A(k, \omega)$, which makes Green's-function-based DMD a natural route toward spectral reconstruction.

## Section S5: Koopman perspective on Green's functions and propagators

A central advantage of Dynamic Mode Decomposition (DMD) is that it can extract effective frequencies, damping rates, and dominant modes directly from numerically-simulated or experimentally-measured time-domain data, without requiring a fully specified microscopic model. This makes it particularly attractive in complex environments involving dissipation, disorder, or moderate nonlinearity, where the dominant linear response remains observable but the full Hamiltonian is difficult to reconstruct. At the same time, its limitations should be kept explicit: the number and quality of extracted modes are constrained by data resolution and noise; strong nonlinearity or strong many-body scattering may broaden the spectrum into a continuum and obscure any clean modal decomposition; and certain topological quantities require global information that cannot be obtained from a single local measurement without additional scans or theoretical input.

For many-body applications, the appropriate input data may take different forms. At the mean-field level, one may analyze time-dependent effective fields. For small finite systems, one may in principle analyze full many-body state vectors, although the Hilbert-space dimension grows exponentially. In this case, the full many-body wavefunction, or alternatively the density matrix, can be regarded as the state of a finite-dimensional dynamical system. Koopman-DMD then approximates the spectral content of the many-body propagator or, at the density-matrix level, the Liouvillian evolution within the chosen finite or truncated Hilbert space. This route is therefore conceptually direct but practically limited to small systems, truncated Hilbert spaces, or reduced dynamical sectors. More generally, one may apply DMD to Green's function, in which case the extracted modes represent quasiparticle-like poles, collective excitations, or relaxation channels.

When many-body interactions become important, the contrast between Hamiltonian Floquet–Bloch methods and Koopman–DMD becomes more pronounced. In conventional Hamiltonian treatments, one usually introduces explicit approximations, such as Hartree–Fock or other mean-field decouplings, in order to reduce quartic interaction terms ($c^\dagger c^\dagger cc$) to an effective single-particle form. By contrast, Koopman–DMD acts directly on higher-order observables, such as density–density or current–current correlation functions, and extracts the dominant coherent structures present in the measured data. From this perspective, DMD may be viewed as a data-driven reduced description of interacting dynamics: the leading modes correspond to effective collective excitations, while strongly incoherent fluctuations are filtered into broad or weakly resolved components. This does not replace controlled many-body theory, but it can provide a compact summary of renormalized quasiparticle-like or collective spectral features when such features remain reasonably well defined.

For interacting many-body systems, a simple single-particle Hamiltonian is generally no longer sufficient. Koopman–DMD instead projects the dynamics onto a chosen observable sector, for example density correlations $O = \rho_i \rho_j$, current correlations, or spin correlations, and analyzes the resulting linearized evolution in observable space. The extracted modes then correspond naturally to collective excitations such as phonons, plasmons, magnons, or order-parameter oscillations. In this way, the method provides a data-driven reduced description of the excitation picture in the chosen observable sector, rather than a full microscopic solution of the many-body problem. Its applicability is clearest in weak-to-intermediate interaction regimes, where identifiable quasiparticle or collective poles still exist. In strongly correlated regimes, however, where the spectrum becomes diffuse and quasiparticles may disappear altogether, DMD is more likely to return decaying relaxation channels than sharply defined elementary excitations.

This limitation is particularly relevant in strongly correlated topological systems, where Hamiltonian methods typically require either many-body topological invariants or Green-function-based formulations of topology, since the usual Bloch-band picture is no longer sufficient. Koopman–DMD in this regime remains exploratory. It may still be useful for analyzing collective modes or boundary-sensitive correlators, but its interpretation is necessarily more indirect than in single-particle settings. Accordingly, Koopman–DMD is currently best viewed as a complementary diagnostic in strongly correlated problems, rather than a substitute for full many-body topological analysis.

The relation between Green's functions and spectral functions provides the natural bridge between these two viewpoints. Since the spectral function is given by $A(\omega) = -\frac{1}{\pi}\,\mathrm{Im}\,G^R(\omega)$, complete spectral information determines the retarded Green's function through the Kramers–Kronig relations. In principle, if the spectral function is measured with sufficient resolution, one may reconstruct an effective Green's function and thereby constrain an effective self-energy or reduced model. In practice, the inverse problem is ill-conditioned and requires smoothing, regularization, and physical priors. Within this context, Koopman–DMD may be regarded as a structured spectral-analysis tool: it converts time-domain signals into a discrete set of damped oscillatory modes, from which one obtains a finite-dimensional approximation to the spectral response. Combined with Hamiltonian modeling or Green-function constraints, this may help guide reduced modeling of partially known systems from experimental data.

We now formulate this point more explicitly from the Koopman perspective. In many-body theory, the natural observables are two-time retarded Green's functions, defined in the Heisenberg picture by

$$G_r(t,t') = -\frac{i}{\hbar}\theta(t-t')\langle[A(t),B(t')]_\pm\rangle \equiv \langle\langle A(t)|B(t')\rangle\rangle_r \tag{S46}$$

where $[\cdot,\cdot]_\pm$ denotes either a commutator or anticommutator, depending on the bosonic or fermionic statistics. In equilibrium, the expectation value is taken with respect to the thermal density matrix. Using the Heisenberg equation of motion $i\hbar\,\frac{dA(t)}{dt} = [A(t),H]$, one obtains the governing equation

$$i\frac{d}{dt}\langle\langle A(t)|B\rangle\rangle_\mathrm{r} = \delta(t)\langle[A(t),B]_\pm\rangle + \langle\langle[A(t),H]|B\rangle\rangle_r. \tag{S47}$$

Note that we set $t' = 0$ for simplicity. After Fourier transformation, this becomes $\langle\langle A|B\rangle\rangle_\omega = \langle[A,B]_\pm\rangle + \langle\langle[A,H]|B\rangle\rangle_\omega$, which immediately generates an infinite hierarchy through repeated commutators,

$$A_1 = [A,H], A_2 = [A_1,H], \ldots, A_{n+1} = [A_n,H]. \tag{S48}$$

From the Koopman–DMD perspective, this hierarchy expresses the difficulty of projected many-body dynamics: the evolution is linear in observable space, but it is not closed in any finite set of observables unless one introduces a closure scheme. DMD may then be interpreted as a **data-driven low-rank closure** of this hierarchy:

$$\begin{aligned}
\omega\langle\langle A|B\rangle\rangle_\omega &= \langle [A,B]_\pm\rangle + \langle\langle A_1|B\rangle\rangle_\omega,\\
\omega\langle\langle A_1|B\rangle\rangle_\omega &= \langle [A_1,B]_\pm\rangle + \langle\langle A_2|B\rangle\rangle_\omega,\\
&\vdots\\
\omega\langle\langle A_{n-1}|B\rangle\rangle_\omega &= \langle [A_{n-1},B]_\pm\rangle + \langle\langle A_n|B\rangle\rangle_\omega.
\end{aligned} \tag{S49}$$

The operators A and B may be either field operators or physical observables. In the first case, one obtains single-particle propagators such as $G_r(t) = -i\theta(t)\langle\psi(t)|\psi^\dagger(0)\rangle$, while in the second case one obtains observable correlators such as

$$G_r(t) = -i\theta(t)\langle\rho_q(t)|\rho_q^\dagger(0)\rangle, G_r(t) = -i\theta(t)\langle j_q(t)|j_q^\dagger(0)\rangle, G_r^{\alpha\beta}(t) = -i\theta(t)\langle S_\alpha(t)|S_\beta(0)\rangle$$
(S50)

In the first case, the relevant poles are associated with single-particle excitations; in the second, they are associated with collective modes or susceptibilities.

As a concrete example, consider an interacting electron system with $H = H_0 + H_{\text{int}}, H_0 = \sum_{k,\sigma} E_k c_{k\sigma}^\dagger c_{k\sigma}$. For the retarded single-particle Green's function, the equation of motion gives

$$\omega\left\langle\langle c_{k\sigma}|c_{k\sigma}^\dagger\rangle\right\rangle_\omega = \left\langle[c_{k\sigma}, c_{k\sigma}^\dagger]_\pm\right\rangle + \left\langle\langle [c_{k\sigma}, H_0]|c_{k\sigma}^\dagger\rangle\right\rangle_\omega + \left\langle\langle [c_{k\sigma}, H_{\text{int}}]|c_{k\sigma}^\dagger\rangle\right\rangle_\omega \tag{S51}$$

Using $[c_{k\sigma}, c_{k\sigma}^\dagger]_\pm = 1, [c_{k\sigma}, H_0] = E_k c_{k\sigma}$, and introducing the retarded self-energy $\Sigma_r(k,\omega)$ to summarize interaction effects ($\left\langle\langle [c_{k\sigma}, H_{\text{int}}]|c_{k\sigma}^\dagger\rangle\right\rangle_\omega \approx \Sigma_r(k,\omega)\left\langle\langle c_{k\sigma}|c_{k\sigma}^\dagger\rangle\right\rangle_\omega$), one obtains the standard Dyson form

$$G_r(k,\omega) = \left\langle\langle c_{k\sigma}|c_{k\sigma}^\dagger\rangle\right\rangle_\omega = \frac{1}{\omega - E_k - \Sigma_r(k,\omega)}. \tag{S52}$$

The associated spectral function is

$$A(\boldsymbol{k},\omega) = -\frac{1}{\pi}\text{Im}G_r(k,\omega) = -\frac{1}{\pi}\frac{\text{Im}\Sigma_r}{[\omega+\mu-E_{\boldsymbol{k}}-\text{Re}\Sigma_r]^2+[\text{Im}\Sigma_r]^2}. \tag{S53}$$

Under the usual quasiparticle-pole approximation, this has the form of a Lorentzian resonance centered at the renormalized energy $\tilde{E}_{\boldsymbol{k}} = E_{\boldsymbol{k}} + \text{Re}\Sigma_r$, with linewidth $\Gamma_k = -\text{Im}\Sigma_r$. In the time domain, the corresponding retarded propagator behaves approximately as

$$G_r(k,t) = \frac{1}{2\pi}\int d\omega \frac{e^{-i\omega t}}{\omega - \tilde{E}_k + i\Gamma_k} = -\frac{i}{\hbar}\theta(t)e^{-\frac{i\tilde{E}_k t}{\hbar}}e^{-\Gamma_k t}, \tag{S54}$$

so that the DMD eigenvalues naturally encode both the renormalized oscillation frequency and the damping rate. Physically, this means that the extracted DMD mode represents a quasiparticle-like excitation that gradually loses coherence because of many-body scattering.

The same logic extends to two-particle Green's functions and susceptibilities. Their poles correspond not to single-particle quasiparticles, but to damped collective oscillations with characteristic frequencies and finite lifetimes. In practice, however, one rarely has access to the full hierarchy of higher-order Green's functions. This is precisely where delay embedding becomes useful. Rather than explicitly constructing all higher-order correlators, one may embed the measured time series of a lower-order Green's function into a higher-dimensional delay-coordinate space using a Hankel matrix,

$$\mathrm{H}_0 = \begin{bmatrix} G_r(t_0) & G_r(t_1) & \cdots & G_r(t_{m-d-1}) \\ G_r(t_1) & G_r(t_2) & \cdots & G_r(t_{m-d}) \\ \vdots & \vdots & \ddots & \vdots \\ G_r(t_{d-1}) & \hat{O}_q & \cdots & G_r(t_{m-1}) \end{bmatrix}, \mathrm{H}_1 = \begin{bmatrix} G_r(t_1) & G_r(t_2) & \cdots & G_r(t_{m-d}) \\ G_r(t_2) & G_r(t_3) & \cdots & G_r(t_{m-d+1}) \\ \vdots & \vdots & \ddots & \vdots \\ G_r(t_d) & G_r(t_{d+1}) & \cdots & G_r(t_m) \end{bmatrix} \tag{S55}$$

Here m is the number of snapshots and d is the embedding dimension. Choosing d sufficiently large allows the delay coordinates to capture the relevant oscillation and memory scales of the data. Physically, this embedding reintroduces hidden degrees of freedom through temporal memory. The fitted linear model,

$$H_1 = KH_0, \tag{S56}$$

may then be interpreted as a finite-dimensional approximation to the projected Koopman dynamics of the Green's function. In this sense, delay-embedded DMD should not be viewed as fitting a bare mean-field Hamiltonian. Rather, it provides a reduced-order approximation to an effective non-Hermitian generator in which memory effects and self-energy-like corrections may be absorbed into the fitted dynamics. This suggests that delay-embedded DMD may offer a useful approximation to the projected Green-function hierarchy at the level of reduced data-driven modeling, although a systematic numerical validation of this interpretation lies beyond the scope of the present work.